\documentclass[fleqn,usenatbib]{mnras}
\usepackage{graphicx}
\usepackage{amsmath}
\usepackage{amssymb}
\usepackage{multicol}
\usepackage{bm}
\usepackage{pdflscape}
\usepackage[T1]{fontenc}
\usepackage{ae,aecompl}
\usepackage{xcolor}

\title[Non-linear theory of the entrainment]{\textit{The entrainment matrix of a superfluid nucleon mixture at finite temperatures}}
\author[Lev B. Leinson]{Lev B. Leinson$^{1}$\thanks{E-mail: leinson@yandex.ru} \\
$^{1}$Pushkov Institute of Terrestrial Magnetism, Ionosphere and Radiowave
Propagation of the Russian Academy of Science (IZMIRAN), \\
108840 Troitsk, Moscow, Russia}
\date{Accepted XXX. Received YYY; in original form ZZZ}
 \pubyear{2018}

\begin{document}
\label{firstpage}
\pagerange{\pageref{firstpage}--\pageref{lastpage}}
\maketitle 

\begin{abstract} 
It is considered a closed system of non-linear equations for the  entrainment matrix of a non-relativistic mixture of superfluid nucleons at arbitrary temperatures below the onset of neutron superfluidity, which takes into account the essential dependence of the superfluid energy gap in the nucleon spectra on the velocities of superfluid flows. It is assumed that the protons condense into the isotropic $^1$S$_0$ state, and the neutrons are paired into the spin-triplet $^3$P$_2$ state. It is derived an analytic solution to the non-linear equations for the entrainment matrix under temperatures just below the critical value for the neutron superfluidity onset. In general case of an arbitrary temperature of the superfluid mixture the non-linear equations are solved numerically  and fitted by simple formulas convenient for a practical use with an arbitrary set of the Landau parameters.
\end{abstract}


\section{Introduction}


It is usually assumed that neutron stars (NSs) are composed mainly of superfluid
nucleons \citep[see, e.g.][]{Takatsuka,syhhp11}. After the pioneering work by \citet{Tamagaki}, it is adopted to assume that the superfluid protons are paired into the spin singlet  $^1$S$_0$, and the superfluid neutrons are in the spin-triplet 
$^3$P$_2$  state in the NS core.
Observations of pulsating NSs, which make it
possible to obtain a unique information on the properties of the superdense
matter, are of a great interest \citep[see, e.g.][]{aks98,a98,fm98,acp02,ak01,gck14,aetal03,gyg05,spa10,pr02}. 
The hydrodynamic theory describing global pulsations of superfluid NSs must necessarily involve the non-dissipative coupling between the nucleons in the NS interior known as the entrainment effect.
Another phenomenon that is also influenced by the entrainment is the
post-glitch response of the NSs. \footnote{As proposed by \citet{ai75}, the
NS glitches are caused by sudden unpinning of a group of vortices
from their pinning centres, resulting in an abrupt increase of the observed
NS rotation frequency.} The core of the star, consisting of the neutron superfluid and plasma of electrons, muons, superconducting protons and, probably, charged and neutral hyperons responds to the glitch via magnetic, viscous and mutual friction forces \citep{als84}.
The mutual entrainment in the superfluid nucleon mixture directly influences the star's response following a glitch via the strong impact on kinetic coefficients of the NS matter, in particular, on the bulk and shear viscosities \citep{ars12,hly00,hly01,sy08}.

The entrainment effect was included in the rotational dynamics of pulsars and in the hydrodynamics of NS pulsations by many authors within the framework of the Fermi liquid theory \citep[see, e.g.][]{pr02,pca02,pr04,ga06,chg11}. Usually the mutual entrainment of supercurrents is described with the aid of the entrainment matrix first introduced by Andreev and Bashkin \citep{ab75} in the hydrodynamics of the superfluid mixture $^3$He and $^4$He.
For the baryon matter of a NS core the most interest represents the entrainment matrix which relates the mass current density of the baryons with the relative velocity of their superfluid and normal components.

For the case of zero temperature the entrainment matrix of the superfluid neutron-proton mixture was calculated  in \citet{Betal96,cj03,ch06,gkh09}. These calculations assume that the energy gaps of superfluid nucleons are constant and, because of a strong dependence of the energy gaps on superfluid velocities, are valid only for small amplitudes of the neutron star's pulsations. Namely, the superfluid velocities of the neutron and proton flows relative the normal (non-superfluid) nucleons should be small as compared to the critical flow velocity at which the superfluidity is destroyed. The formalism for calculations of the entrainment matrix at a zero temperature valid for arbitrary superfluid velocities was developed by \citet{Leinson2017}.

Obviously, the zero-temperature theory can be applied in many cases, since most of isolated objects older than $10^5$ years have the temperature much below the critical temperature for the nucleon superfluidity onset \citep[see, e.g.][]{yp04,bcf17}. However, there are many cases when the temperature dependence of the entrainment matrix is very important, for example, in pulsations of warm neutron stars, with the temperature of the order of the critical temperature for the neutron (or proton) superfluidity. The temperature dependence can be also important when the NS pulsation energy is higher than its thermal energy and the star can heat by the pulsation energy dissipation \citep{gyg05}. As the mutual entrainment in the superfluid nucleon mixture has a strong impact on  the bulk and shear viscosities its temperature dependence is crucial  in the study of the kinetics of NS.
In these cases the conditions for the unchanged energy gaps become stronger restricted. Namely, the superfluid velocities of the neutron
and proton ($a=n,p$) flows relative the normal (non-superfluid) nucleons are restricted
by the condition $v_{a}\ll v_{a}^{max}(T)$, where $v_{a}^{max}(T)\sim\Delta
^{\left( a\right)}\left( T\right) /k_{Fa}$ is the critical speed of the
superfluid flow at which the superfluidity is destroyed, $\Delta ^{\left(
a\right)}\left(T\right)$ is the temperature-dependent energy gap for
Bogoliubov excitations in the superfluid at rest, and $k_{Fa}$  is the Fermi momentum of the nucleon specie $a$ (We use natural units $c=\hbar =1$, and 
$k_{\mathrm{B}}=1$.). This condition particularly restricts the theory at the temperatures $T\sim T_{ca}$, where the superfluid energy gap is reduced in comparison with its value at zero temperature.

The aim of this paper is to develop a non-linear theory of the non-relativistic entrainment matrix accounting for the gap dependence on the superfluid velocities  and the temperature. The current work is an extension of the special cases of \citet{Leinson2017} and \citet{gh05}.

The paper is organized as follows. In Sec. \ref{sec:be}, we derive the general form of the non-linear equations for calculating the entrainment matrix of the superfluid nucleon mixture. Further we discuss the gap equations for the spin-singlet pairing of protons and for the spin-triplet pairing of neutrons in a moving superfluid.
In Sec. \ref{sec:fp}, we specify typical values of the physical system parameters under consideration. Sec. \ref{sec:mix} contains the solution to the non-linear equations for the entrainment matrix under the temperature regime typical for a superfluid core of a NS. Section \ref{sec:conc} contains the summary of the obtained results. Appendixes A and B contain some additional information on the spin-triplet pairing of neutrons.

\section{Basic equations}
\label{sec:be}
For simplicity let us consider the mixture of superfluid neutrons and protons which contains also normal electrons and, probably, muons providing the electric quasi-neutrality of the system. In the non-relativistic system, the symmetric entrainment
matrix $\rho_{ab}$ $\left(a,b =n,p\right)$ can be defined by the
relation: 
\begin{align}
\mathbf{j}_{n}& =\left( m_{n}n_{n}-\rho _{nn}-\rho _{np}\right) \boldsymbol{v}%
_{q}+\rho _{nn}\boldsymbol{v}_{n}+\rho _{np}\boldsymbol{v}_{p},  \label{jjn} \\
\mathbf{j}_{p}& =\left( m_{p}n_{p}-\rho _{pp}-\rho _{pn}\right) \boldsymbol{v}%
_{q}+\rho _{pp}\boldsymbol{v}_{p}+\rho _{pn}\boldsymbol{v}_{n},  \label{jjp}
\end{align}%
where $m_{a}$ and $n_{a}$ are the nucleon mass and number density of nucleon
species $a$, respectively; $\mathbf{j}_{a}$ and $\boldsymbol{v}_{a}$ denote the
mass current densities and the superfluid velocities of the neutron and proton components. To avoid confusion notice that the phenomenological superfluid
velocities $\boldsymbol{v}_{a}$ do not represent the nucleon velocities %
\citep[see][]{pr04,ch06}. From Eqs. (\ref{jjn}), (\ref{jjp}) it is apparent
that in general the mass currents are not aligned with the respective superfluid
velocities. We will see below that the kinematic quantities directly
proportional to the mass currents are related to the so-called effective
velocity, defined in Eq. (\ref{V}). Finally, $\boldsymbol{v}_{q}$ is the
hydrodynamic velocity of the normal constituent consisting of electrons and
Bogoliubov excitations of the nucleons.  At a finite temperature, excitations
are present in both superfluid systems, in principle moving at different
speeds. However, because of collisions, all these normal particles (bogolons) quickly reach equilibrium acquiring a common hydrodynamic velocity and can be considered comoving in the absence of external fields.

We assume that the charge densities of electrons, muons and protons are strictly balanced, $n_{e}+n_{\mu}=n_{p}$. Since the NS pulsations period is usually much larger than the inverse plasma frequency of electrons and protons in npe matter \citep[see, e.g.][]{pr02,gyg05} we neglect the electrodynamic effects and assume the electrons and protons to be strictly moving together. 

From Eqs. (\ref{jjn}), (\ref{jjp}) it follows that the entrainment matrix
connects the mass current densities and the velocities of the superfluid
flows of nucleons relative to their normal components 
$\boldsymbol{v}_{a}-\boldsymbol{v}_{q}$. For the sake of
simplicity we consider the mixture of superfluids in the comoving coordinate
frame, where $\boldsymbol{v}_{q}=0$. In this case the superfluid velocities
$\boldsymbol{v}_{a}$ coincide with their relative velocities.

The mutual entrainment in the mixture of superfluids is caused by the
Fermi-liquid interactions. Since the uniform motion of a unitary superfluid
liquid induces no spin polarization we shall use the spin-averaged
Fermi-liquid interactions, which are parametrized by the functions $%
f^{ab}\left( \mathbf{k,k}^{\prime}\right) $. In this case a summation over
the spin projections $\sigma =\uparrow ,\downarrow $ leads to a simple 
multiplication by a factor
of two. Assuming that the velocities of superfluid flows are small in a
scale of the Fermi velocities one may approximately put the arguments of the
functions $f^{ab}\left( \mathbf{k,k}^{\prime}\right) $ be equal to their
values at the corresponding Fermi surfaces. This allows one to write the
interaction function in the form of expansion in Legendre polynomials
parametrized by the Landau parameters $f_{l}^{ab}$: 
\begin{equation}
f^{ab}\left( \mathbf{k,k}^{\prime}\right) =\sum_{l}f_{l}^{ab}P_{l}\left(
\cos\theta_{\mathbf{kk}^{\prime}}\right) ,~\ \ \cos\theta_{\mathbf{kk}%
^{\prime}}\equiv\mathbf{\hat{k}\hat{k}}^{\prime}.  \label{FLint}
\end{equation}
Hereafter $\mathbf{\hat{k}=k}/k$ means a unit vector in the direction of a
quasi-particle momentum $\mathbf{k}$.

We restrict our consideration to a uniform motion of the superfluid flows of
protons and neutrons in unitary states. In this case a uniform motion of all
or part of the liquid induces no spin polarization, nor any change of the
total density, so the only Landau parameters to come in are the $f_{l}^{ab}$
with $l\geq1$. For a nucleon matter the corresponding Landau parameters were
calculated for various mean-field models in a series of papers %
\citep{matsui81,hm87,cgl01,cgl02,cgl03}, where it is shown that only
first two spin-averaged Landau parameters are non-zero, i.e., 
$f_{l}^{ab}=0$ at $l\geq2$. In view of this observation, one can employ only
the parameter $f_{1}^{ab}$, which satisfies the condition $%
f_{1}^{ab}=f_{1}^{ba}$ \citep{S73,lp80}. For more convenience we define the
standard dimensionless Landau parameters%
\begin{equation}
F_{1}^{ab}\equiv f_{1}^{ab}\sqrt{N_{0a}N_{0b}},  \label{Fab}
\end{equation}
where $N_{0a}=m_{a}^{\ast}k_{Fa}/\pi^{2}$. Hereafter $m_{a}^{\ast}$ is the
effective mass of the nucleon which is defined via the Fermi velocity 
$v_{Fa}=k_{Fa}/m_{a}^{\ast}$.

Since we adopt that the flow velocities are small compared to the Fermi
velocity of the nucleons\footnote{%
It is well known the flow velocity needed to destroy the superfluidity is much
smaller than the Fermi velocity of the degenerate nucleons 
\citep{bardeen62}.} the correction to
the quasi-particle energy caused by the superfluid motion can be written up
to the first order in small parameters $v_{a}/v_{Fa}\ll1$. Then the
quasi-particle energy can be written in the form \ 
\begin{equation}
\tilde{\varepsilon}_{\mathbf{k}}^{(a)}=\xi_{k}^{\left( a\right) }+\sum _{b}%
\mathsf{\tilde{\gamma}}_{ab}\mathbf{k}\boldsymbol{v}_{b},  \label{gam}
\end{equation}
where 
\begin{equation}
\xi_{k}^{\left( a\right) }=v_{Fa}\left( k-k_{Fa}\right)  \label{ksi}
\end{equation}
is the quasi-particle energy in the superfluid at rest,\footnote{%
We omit the renormalization of the quasi-particle energy caused by the
superfluidity in the Fermi liquid at rest. This correction  
$\sim (\Delta /\mu)^2 $  is small and
normally ignored \citep[see, e.g.][]{lp80}.} and $\mathsf{\tilde{\gamma}}_{ab}$ is some unknown
matrix, which is taken at the Fermi surface of particle species (at $%
k=k_{Fb} $). This matrix depends on the velocities $\boldsymbol{v}_{p}$ and 
$\boldsymbol{v}_{n}$ since the energy gaps in the excitation spectra of the
quasi-particles depend on the velocities of the superfluid motion.\footnote{%
In \cite{Leinson2017} this matrix is denoted as $\gamma_{\alpha\beta}$.}
Throughout the text we shall omit the velocity arguments, using instead a
tilde above a letter thus indicating the quantities that depend on the
velocities of the superfluid flows.

In the case of spin-singlet isotropic pairing of protons the form (\ref{gam}%
) is quite general because there are only two vectors $\mathbf{k}$ and $%
\boldsymbol{v}_{p}$ to form a scalar correction to the energy. It will be shown
below that this form is also justified in the case of anisotropic
spin-triplet neutron pairing, since it corresponds to the lowest energy of a
homogeneous superfluid flow at a fixed velocity.

It is convenient to introduce the auxiliary vector functions $\mathbf{%
\tilde {V}}_{a}$, which components are defined as 
\begin{equation}
\mathbf{\tilde{V}}_{a}\equiv\sum_{b}\mathsf{\tilde{\gamma}}_{ab}
\boldsymbol{v}_{b}.  \label{V}
\end{equation}
These functions will be called the effective velocities of the superfluid
flows. With the aid of the effective velocities the quasi-particle energy in
the mixture of moving superfluid Fermi liquids can be written as%
\begin{equation}
\tilde{\varepsilon}_{\mathbf{k}}^{(a)}=\xi_{k}^{\left( a\right) }+\mathbf{k%
\tilde{V}}_{a}.  \label{ksitpm}
\end{equation}
Comparison of this expression with the equation (\ref{ksipm}) tells us that
in order to take into account Fermi-liquid interactions, in the mixture of
superfluids, one only needs to replace the flow velocity $\mathbf{v}_{s}$
in the single-particle Green's function (see Appendix A) by the effective
velocity $\mathbf{\tilde{V}}_{a }$. This allows one to immediately
obtain the distribution functions of nucleon quasi-particles in the mixture
of moving superfluid liquids. For the nucleon constituent "$a$" we get
(Greek letters $\alpha,\beta$ refer to the spin indices):
\begin{equation}
\tilde{n}_{\alpha \beta }^{\left( a\right) }\left( \mathbf{k}\right)
=T\sum_{r=-\infty }^{\infty }\mathcal{\tilde{G}}_{\alpha \beta }^{\left(
a\right) }\left( \omega _{r},\mathbf{k}\right) ,  \label{nab}
\end{equation}%
where the Green function $\tilde{G}_{\alpha \beta }^{\left( a\right) }$ is
as given in Eq. (\ref{tgf}) but with $\mathbf{v}_{s}$ replaced by the
effective velocity $\mathbf{\tilde{V}}_{a }$. Using the formula for
summation over fermion Matsubara frequencies $\omega _{r}=\left(
2r+1\right) \pi T$:%
\begin{equation}
T\sum_{r=-\infty }^{\infty }\frac{1}{i\omega _{r}-x}=\frac{1}{%
1+e^{x/T}}  \label{sor}
\end{equation}%
one can get $\tilde{n}_{\alpha \beta }^{\left( a\right) }\left( \mathbf{k}%
\right) =\delta _{\alpha \beta }\tilde{n}_{\mathbf{k}}^{\left( a\right) }$,
where%
\begin{equation}
\tilde{n}_{\mathbf{k}}^{\left( a\right) }=\mathtt{\tilde{v}}_{\mathbf{k}%
}^{\left( a\right) 2}+\mathtt{\tilde{u}}_{\mathbf{k}}^{\left( a\right) 2}%
\mathcal{\tilde{F}}_{+}^{\left( a\right) }-\mathtt{\tilde{v}}_{\mathbf{k}%
}^{\left( a\right) 2}\mathcal{\tilde{F}}_{-}^{\left( a\right) }.  \label{tn}
\end{equation}
The parameters $\mathtt{\tilde{u}}_{\mathbf{k}}^{\left( a\right) }$ and $%
\mathtt{\tilde{v}}_{\mathbf{k}}^{\left( a\right) }$ are defined as%
\begin{equation}
\mathtt{\tilde{u}}_{\mathbf{k}}^{\left( a\right) 2}=\frac{1}{2}\left( 1+%
\frac{\xi _{k}^{\left( a\right) }}{\tilde{E}_{\mathbf{k}}^{\left( a\right) }}%
\right) ,  \label{uu}
\end{equation}%
\begin{equation}
\mathtt{\tilde{v}}_{\mathbf{k}}^{\left( a\right) 2}=\frac{1}{2}\left( 1-%
\frac{\xi _{k}^{\left( a\right) }}{\tilde{E}_{\mathbf{k}}^{\left( a\right) }}%
\right) ,  \label{vv}
\end{equation}
and the distribution functions for Bogoliubov excitations (bogolons) are
given by 
\begin{equation}
\mathcal{\tilde{F}}_{\pm }^{\left( a\right) }\left( \mathbf{k}\right) =\frac{%
1}{1+e^{\left( \tilde{E}_{\mathbf{k}}^{\left( a\right) }\pm \mathbf{k\tilde{V%
}}_{a}\right) /T}}.  \label{dis}
\end{equation}%
The bogolon energy, 
\begin{equation}
\tilde{E}_{\mathbf{k}}^{\left( a\right) }=\sqrt{\xi _{k}^{\left( a\right) 2}+%
\tilde{D}_{a}^{2}(T,\mathbf{\hat{k}})},  \label{EpD}
\end{equation}%
depends on the effective flow velocity $\mathbf{\tilde{V}}_{a}$ through the
energy gap $\tilde{D}_{a}^{2}(T,\mathbf{\hat{k}})$ (see below).

To derive equations for the unknown matrix $\mathsf{\tilde{\gamma}}_{ab}$
let us write the change of the quasi-particle energy due to the superfluid
motion, which follows from the Fermi liquid theory in the limit $\left\vert 
\boldsymbol{Q}_{a }\right\vert \ll k_{Fa}$
\begin{equation}
\tilde{\varepsilon}_{\mathbf{k}}^{(a)}=\xi _{k}^{\left( a\right) }+\frac{%
\mathbf{kQ}_{a }}{m_{a}^{\ast }}+2\sum_{\mathbf{k}^{\prime }b}f_{1}^{ab}%
\mathbf{\hat{k}\hat{k}}^{\prime }\delta \tilde{n}^{\left( b\right) }\left( 
\mathbf{k}^{\prime }\right) .  \label{epsq}
\end{equation}%
Here $\mathbf{Q}_{a }=m_{a}\boldsymbol{v}_{a}$; the Landau effective
mass accounts for the Fermi liquid
interactions in the superfluid at rest. The factor of $2$ in the last term
arises due to summation over spins.  The change of the distribution
functions because of the supercurrents is given by%
\begin{equation}
\delta \tilde{n}_{\mathbf{k}}^{(a )}\equiv \left( \tilde{n}_{\mathbf{k}%
}^{(a )}-\Theta _{\mathbf{k+Q}_{a}}^{(a )}\right) -\left( n_{%
\mathbf{k}}^{(a )}-\Theta _{\mathbf{k}}^{(a )}\right) ,  \label{dn}
\end{equation}%
where $\Theta _{\mathbf{k}}^{\left( a \right) }\equiv \Theta \left(
k_{Fa }-\left\vert \mathbf{k}\right\vert \right) $~is the step
function, and $n_{\mathbf{k}}^{(a )}$ and $n_{\mathbf{k}}^{(a )}$
stands for the distribution of quasi-particles in the superfluid mixture at
rest. It is given by the same expression (\ref{tn}) but with $\mathbf{\tilde{%
V}}_{a}=0$.  Since the relations $Q_{a }\ll k_{Fa }$, or
equivalently, $v_{a}\ll v_{Fa}$ are well fulfilled, in
superfluid Fermi liquids, one can write $\Theta _{\mathbf{k+Q}_{a}}^{(a
)}-\Theta _{\mathbf{k}}^{(a )}\simeq -m_{a}\mathbf{\tilde{V}}_{a }%
\mathbf{\hat{k}\,}\delta \left( k-k_{Fa }\right) $.

Equations for the unknown components of the matrix 
$\mathsf{\tilde{\gamma}}_{ab}$ can be obtained from a comparison of 
Eqs. (\ref{epsq}) and (\ref{gam}). We get%
\begin{equation}
\sum_{b}\mathsf{\tilde{\gamma}}_{ab}\mathbf{k}\boldsymbol{v}_{b}=\frac{m_{a}%
}{m_{a}^{\ast }}\mathbf{k}\boldsymbol{v}_{a }+2\sum_{b\mathbf{k}%
^{\prime }}f_{1}^{ab}\mathbf{\hat{k}\hat{k}}^{\prime }\delta \tilde{n}%
^{\left( b\right) }\left( \mathbf{k}^{\prime }\right) .  \label{ggeq}
\end{equation}%
In a standard way the summation over $\mathbf{k}^{\prime }$ can be converted
into the integral%
\begin{equation}
2\sum_{\mathbf{k}}\equiv \int \frac{2d^{3}k}{\left( 2\pi \right) ^{3}}\cdot
\cdot \cdot \simeq \frac{k_{F_{a }}m_{a }^{\ast }}{\pi ^{2}}%
\int_{-\infty }^{\infty }d\xi _{k}\int \frac{d\mathbf{\hat{k}}}{4\pi }\cdot
\cdot \cdot .  \label{sigP}
\end{equation}%
The integral with respect to the angles on the right-hand side of Eq. 
(\ref{ggeq}) can be carried out using the addition theorem, which is valid for the
Legendre polynomials:%
\begin{equation}
\mathbf{\hat{k}\hat{k}}^{\prime }=\cos \theta _{\mathbf{kV}}\cos \theta _{%
\mathbf{k}^{\prime }\mathbf{V}}+\sin \theta _{\mathbf{kV}}\sin \theta _{%
\mathbf{k}^{\prime }\mathbf{V}}\cos \left( \phi -\phi ^{\prime }\right) .
\label{leg}
\end{equation}%
After integration we write $\mathbf{\hat{k}\tilde{V}}_{a}=\sum_{b}\mathsf{%
\tilde{\gamma}}_{ab}\mathbf{\hat{k}}\boldsymbol{v}_{b}$ and equate terms
with the same $\mathbf{\hat{k}}\boldsymbol{v}_{a}$ in the left- and
right-hand sides thus obtaining the set of equations for $\mathsf{\tilde{%
\gamma}}_{ab}$:
\begin{align}
\mathsf{\tilde{\gamma}}_{aa} & ={\frac{m_{a}}{m_{a}^{\ast}}\frac{1}{S}}\left[
\left( 1+{\frac{F_{1}^{aa}}{3}}\right) \left( 1+{\frac{F_{1}^{bb}}{3}}\,%
\tilde{\Phi}_{b}\right) -{\frac{F_{1}^{ab}F_{1}^{ba}}{9}}\tilde{\Phi }_{b}%
\right] ,  \label{s1} \\
\mathsf{\tilde{\gamma}}_{ab} & ={\frac{1}{3}}\,\,{\frac{m_{b}}{\sqrt {%
m_{a}^{\ast}\,m_{b}^{\ast}}}}\,\,{\frac{1}{S}}\,\,\left( {\frac{k_{Fb}}{%
k_{Fa}}}\right) ^{3/2}\,\,F_{1}^{ab}\,(1-\tilde{\Phi}_{b})\,,  \label{s2} \\
S & \equiv\left( 1+{\frac{F_{1}^{aa}}{3}}\,\tilde{\Phi}_{a}\right) \,\left(
1+{\frac{F_{1}^{bb}}{3}}\,\tilde{\Phi}_{b}\right) -{\frac {%
F_{1}^{ab}F_{1}^{ba}}{9}}\,\tilde{\Phi}_{a}\tilde{\Phi}_{b}.  \label{S}
\end{align}
Here $a\neq b$ so that if $a=n$ then $b=p$ and vice-versa. The functions $%
\tilde{\Phi}_{a}$ are defined by the integral 
\begin{equation}
\tilde{\Phi}_{a}\equiv-\frac{3}{N_{0a}}\frac{1}{k_{Fa}\tilde{V}_{a}}\int%
\frac{d^{3}k}{8\pi^{3}}\left( \mathcal{\tilde{F}}_{\mathbf{+}}^{(a)}-%
\mathcal{\tilde{F}}_{-}^{(a)}\right) \cos\theta_{\mathbf{k\tilde{V}}%
_{a}}.  \label{Fi}
\end{equation}

The equations must be supplemented by an expression for the effective mass
which follows from Galilean invariance \citep{S73,Betal96}: 
\begin{equation}
{\frac{m_{a}^{\ast }}{m_{a}}}\,\,=1+\frac{N_{0a }}{3}\left[ f_{1}^{aa}+%
\frac{m_{b}}{m_{a }}\left( \frac{k_{Fb}}{k_{Fa }}\right)
^{2}f_{1}^{ab}\right] .  \label{mef}
\end{equation}%
It is convenient to recast this relation in terms of the dimensionless
Landau parameters separately for protons and neutrons:%
\begin{equation}
{\frac{m_{p}^{\ast }}{m_{p}}}\,\,=1+\frac{1}{3}F_{1}^{pp}+\frac{1}{3}\sqrt{%
\frac{m_{p}^{\ast }}{m_{n}^{\ast }}}\frac{m_{n}}{m_{p}}\left( \frac{k_{Fn}}{%
k_{Fp}}\right) ^{3/2}F_{1}^{pn},  \label{mefp}
\end{equation}%
\begin{equation}
{\frac{m_{n}^{\ast }}{m_{n}}}\,\,=1+\frac{1}{3}F_{1}^{nn}+\frac{1}{3}\sqrt{%
\frac{m_{n}^{\ast }}{m_{p}^{\ast }}}\frac{m_{p}}{m_{n}}\left( \frac{k_{Fp}}{%
k_{Fn}}\right) ^{3/2}F_{1}^{np}.  \label{mefn}
\end{equation}%
The system of equations (\ref{mefp}) and (\ref{mefn}) along with the $%
F_{1}^{np}=F_{1}^{pn}$ symmetry ratio makes it possible to calculate the
effective masses of protons and neutrons for a given set of dimensionless
Landau parameters.

Notice, the effective velocity $\mathbf{\tilde{V}}_{a}$, as defined by Eq. (%
\ref{V}), substantially depends on the unknown matrix $\mathsf{\tilde{\gamma 
}}_{ab}$, so that the Eqs. (\ref{s1})-(\ref{S}) are highly non-linear. If
the superfluid velocities are small in comparison with the critical
velocities necessary for the destruction of superfluidity, $%
v_{a}\ll\Delta^{\left( a\right) }\left( T\right) /k_{Fa}$, one may
neglect the dependence on the velocities in the functions (\ref{Fi}) and
adopt the gap amplitude be constant. In this limit, Eqs. (\ref{s1})-(\ref{S}%
) recover the result obtained in \citet{gh05}.

Following the Fermi-liquid theory, the mass current density can be evaluated
from the same expression, which is normally used in the case of
non-superfluid matter \citep[see][]{leg65,leg75} 
\begin{equation}
\mathbf{j}_{a}=\sum_{\mathbf{k,}\sigma }m_{a}\frac{\partial \tilde{%
\varepsilon}_{\mathbf{k}}^{(a)}}{\partial \mathbf{k}}\tilde{n}_{\mathbf{k}%
}^{(a)}.  \label{J}
\end{equation}%

Substituting Eqs. (\ref{gam}) and (\ref{tn}) into Eq. (\ref{J}) and
performing simple integrations we find
\begin{equation}
\mathbf{j}_{a}=m_{a}n_{a}\mathbf{\tilde{V}}_{a}(1-\tilde{\Phi}_{a}),  
\label{JFi}
\end{equation}%
where the functions $\tilde{\Phi}_{a }$ is defined in Eq. (\ref{Fi}).

Having obtained the matrix $\mathsf{\tilde{\gamma}}_{ab}$ one can write the
entrainment matrix for arbitrary velocities of the superfluid flows %
\citep{Leinson2017}%
\begin{equation}
{\rho }_{a b}=\rho _{a }\,\mathsf{\tilde{\gamma}}_{a
b}\,\left( 1-\tilde{\Phi}_{a }\right) ,  \label{rab}
\end{equation}%
where $\rho _{a }=m_{a }n_{a }$. It represents a complicated
non-linear function of the velocities of superfluid flows $\boldsymbol{v}_{p}
$ and $\boldsymbol{v}_{n}$.

The Eqs. (\ref{s1})-(\ref{S}) should be solved simultaneously with the gap
equations in order to consistently take into account the gap dependence on
the flow velocities and temperature. It is thought that pairing of neutrons in superdense
core of NSs occurs into the $^{3}$P$_{2}$ state (with a small
admixture of $^{3}$F$_{2}$) while other baryons undergo the spin-singlet
pairing $^{1}$S$_{0}$ below the corresponding critical temperature $T_{ca}$
for the superfluidity onset. Suppose we know the amplitude of the gap in the
superfluid liquid, which is at rest at zero temperature. Our goal is to find
the dependence of the energy gap on the velocities of the superfluid flows
at different temperatures. This problem has been repeatedly considered for
the superfluid flow in the BCS approximation neglecting the Fermi-liquid
effects \citep{bardeen62,alex03,gk13,ft72,Leinson2017}.

It is easy to see from Eqs. (\ref{s1})-(\ref{S}) that $\mathsf{\tilde{\gamma}%
}_{ab}\rightarrow \delta _{ab}m_{a}/m_{a}^{\ast }$, if we turn off the
Fermi-liquid interactions, by putting formally $F_{1}^{aa}=F_{1}^{ab}=0$. In
this case from Eq. (\ref{V}) we find $\mathbf{\tilde{V}}_{a
}\rightarrow \left( m_{a}/m_{a}^{\ast }\right) \boldsymbol{v}_{a}$, and the
distribution functions, given by Eq. (\ref{dis}), transform to the standard
distribution functions for bogolons in a free nucleon gas with pairing. This
observation tells us that the effects of a Fermi liquid, in the superfluid
mixture of nucleons, will be taken into account if, in the gap equations
obtained in the BCS theory, we replace the velocity $\mathbf{v}_{a}$ of a
superfluid flow of one type of nucleons onto the effective velocity $\mathbf{%
\tilde{V}}_{a }$, which is defined in Eq. (\ref{V}).

\subsection{Spin-singlet pairing of protons in a moving condensate}
\label{sec:sp}
Consider, for instance, the spin-singlet pairing of protons in the nucleon
liquid. The BCS equation for the gap amplitude in the moving $^{1}$S$_{0}$
superfluid at temperature $T$, is well known. See e.g. Eq. (B7) in %
\citet{bardeen62}. From the above discussion it follows that the Fermi
liquid interactions can be taken into account with the aid of the
replacement $\mathbf{v}_{p}\rightarrow \mathbf{\tilde{V}}_{p}$ in this
equation. Combining the obtained equation for the gap amplitude $\tilde{%
\Delta}^{\left( p\right) }$ in the superfluid mixture with supercurrents at
temperature $T$ with the same equation for the immovable superfluid at $T=0$
one can obtain [compare with \citet{alex03,gk13}]
\begin{align}
\frac{k_{Fp}m_{p}^{\ast }}{\pi ^{2}}\ln \frac{\Delta _{0}^{\left( p\right) }%
}{\tilde{\Delta}^{\left( p\right) }}& =\sum_{\mathbf{k}}\frac{1}{\tilde{E}_{%
\mathbf{k}}^{\left( p\right) }}\left[ \frac{1}{1+e^{\left( \tilde{E}_{%
\mathbf{k}}^{\left( p\right) }+\mathbf{k\tilde{V}}_{p}\right) /T}}\right.  
\notag \\
& \left. +\frac{1}{1+e^{\left( \tilde{E}_{\mathbf{k}}^{\left( p\right) }-%
\mathbf{k\tilde{V}}_{p}\right) /T}}\right] ,  \label{sing}
\end{align}%
where the bogolon energy 
\begin{equation}
\tilde{E}_{\mathbf{k}}^{\left( p\right) }=\sqrt{\xi _{k}^{\left( p\right) 2}+%
\tilde{\Delta}^{\left( p\right) 2}}  \label{EP}
\end{equation}%
depends on the effective flow velocity $\mathbf{\tilde{V}}_{p}$ through the
isotropic energy gap $\tilde{\Delta}^{\left( p\right) }$. 
We denote $\Delta _{0}^{\left(p\right) }$ the energy gap amplitude in the superfluid at rest and
temperature $T=0$. In the case of $^{1}$S$_{0}$ pairing of protons this
value is known to be%
\begin{equation}
\Delta _{0}^{\left( p\right) }=\pi e^{-C}T_{cp}=1.\,\allowbreak 764T_{cp},
\label{dp0}
\end{equation}%
where $C=0.577216$ is Euler's constant.

\subsection{Spin-triplet pairing of neutrons in a moving condensate}
\label{sec:tp} 
For the spin-triplet pairing the equation that allows one to find the gap amplitude $\tilde{\Delta }%
^{\left( n\right) }$ as a function of the temperature and effective velocity 
\textbf{$\tilde{V}$}$_{n}$, in the superfluid mixture, can be written in the
form \citep{ft72,Leinson2017}:
\begin{equation}
\frac{k_{Fn}m_{n}^{\ast}}{\pi^{2}}\ln\frac{\Delta_{0}^{\left( n\right) }}{%
\tilde{\Delta}^{\left( n\right) }}=\sum_{\mathbf{k}}\frac{\bar{b}^{2}(%
\mathbf{\hat{k}})}{\tilde{E}_{\mathbf{k}}^{\left( n\right) }}\left( \mathcal{%
\tilde{F}}_{+}^{\left( n\right) }+\mathcal{\tilde{F}}_{-}^{\left( n\right)
}\right) .  \label{trip}
\end{equation}
In this equation, $\Delta_{0}^{\left( n\right) }$ is the energy gap
amplitude in the superfluid at rest (i.e. for \textbf{$\tilde{V}$}$_{n}=0$)
and temperature $T=0$. The distribution functions for the Bogoliubov
excitations $\mathcal{\tilde{F}}_{\pm}^{\left( n\right) }$ are as defined in
Eq. (\ref{dis}) with%
\begin{equation}
\tilde{E}_{\mathbf{k}}^{\left( n\right) }=\sqrt{\xi_{k}^{\left( n\right) 2}+%
\tilde{\Delta}^{\left( n\right) 2}\bar{b}^{2}(\mathbf{\hat{k}})}.  \label{Ep}
\end{equation}
The real vector $\mathbf{\bar{b}}(\mathbf{\hat{k}})$ defines the angle anisotropy
of the energy gap which depends on the phase state of the superfluid condensate (see Appendix B).
Note that if one formally puts $\bar{b}\rightarrow1$, the equation (\ref%
{trip}) becomes identical to Eq. (\ref{sing}).

\section{Formulation of the problem}
\label{sec:fp} 
Apparently, the system of non-linear equations for the
entrainment matrix can not be solved in general form. To proceed one needs
to specify typical values of the physical system parameters under
consideration.

Although microscopic calculations are model
dependent, it is customary to assume that the critical temperature for a
superfluidity onset typical for $^{1}$S$_{0}$ pairing of protons or hyperons
are about an order of magnitude higher than the critical temperature for the 
$^{3}$P$_{2}$ pairing of neutrons in the NS core %
\citep[see, e.g.][]{tnyt01,ws10,ccdk91,eeho96,tt04}. Thus at a temperature $T\lesssim T_{{%
cn}}$, when the superfluid liquid of neutrons should be considered warm, the
superfluid protons and hyperons are cold, $T\ll T_{{ca}}$ ($a=p,\Lambda
,\Sigma ^{-}$) and can be considered in the limit of zero temperature. Let
us adopt this assumption, which substantially simplifies the problem.

Another important observation is that the neutron Fermi momentum is the largest among the baryon constituents of the NS core. Since the critical
superfluid flow velocity necessary to destroy the superfluidity of the
baryon specie "$a$" ($a=n,p,\Lambda ,\Sigma ^{-}$) is of the order of 
\begin{equation}
v_{a}^{\max }\sim \frac{T_{ca}}{k_{Fa}},  \label{vma}
\end{equation}%
the superfluid velocity necessary to destroy the spin-triplet pairing of
neutrons is small as compared to the critical velocity destroying the
spin-singlet pairing of protons or hyperons. 

If we assume that all the components of the baryon mixture participating in the superfluid motion have superfluid velocities of the same order of magnitude, we can conclude that only the neutron energy gap will be sensitive to superfluid motions. 
On the contrary, the change in the energy
gaps caused by the spin-singlet pairing of protons and hyperons should be
negligibly small, since the existence of neutron superfluidity indicates
that the velocities of the superfluid flows are small in comparison with the
critical velocities for pairing of protons or hyperons. 

From the above discussion it follows that generalization to the case of the baryon matter involving hyperons is straightforward. Therefore we restrict our analysis to the case of $npe\mu$ matter. 

The equations (\ref{s1})-(\ref{Fi}) assume that the superfluid velocities of
neutrons and protons are independent parameters of the problem. This,
however, is not the case in actual applications. Consider, for example, the
motion of superfluid nucleon mixture in the inner core of pulsating neutron
star. Suppose that the normal (not superfluid) component of the nucleon
matter consists of nucleon excitations, electrons and, probably, muons. In
the absence of external fields, collisions between these particles lead to
the fact that they have the same hydrodynamic velocity $\boldsymbol{v}_{q}$.
This velocity is equal to zero in the comoving coordinate system, which we
use. This means that the electric currents in the comoving system can be
written as

\begin{equation}
\mathbf{j}_{p}=\rho _{pp}\boldsymbol{v}_{p}+\rho _{pn}\boldsymbol{v}_{n},~%
\mathbf{j}_{e}=\mathbf{j}_{\mu }=0.  \label{jpe}
\end{equation}%
The pulsations period is usually much greater than the inverse plasma frequency 
\citep[see, e.g.][]{ga06}. In this case, only longitudinal oscillations are possible in a quasi-neutral liquid, and the quasi-neutrality condition must hold. 
In the case of $npe\mu$ matter this means
\begin{equation}
n_{p}=n_{e}+n_{\mu }.  \label{qnc}
\end{equation}%
Then from the particle conservation laws ($l=n,p,e,\mu $),%
\begin{equation}
\partial _{t}n_{l}+\boldsymbol{\nabla }\mathbf{j}_{l}=0,  \label{pcl}
\end{equation}%
one gets $\boldsymbol{\nabla }\mathbf{j}_{p}=0$. For longitudinal plane waves
of the form $\sim e^{i\left( \omega t-\mathbf{qr}\right) }$, where $%
\omega $ is the perturbation frequency and $\mathbf{q}$ is the wave vector,
this yields 
\begin{equation}
\rho _{pp}\boldsymbol{v}_{p}+\rho _{pn}\boldsymbol{v}_{n}=0\text{.}
\label{rel}
\end{equation}%
This condition is normally used in studying the oscillation spectrum of
pulsating superfluid NSs \citep[see, e.g.][]{m91,ga06,chg11}.

Making use of Eqs. (\ref{V}), (\ref{rab}) and (\ref{rel}) we get%
\begin{equation}
\mathbf{\tilde{V}}_{p}\equiv \mathsf{\tilde{\gamma}}_{pp}\boldsymbol{v}_{p}+%
\mathsf{\tilde{\gamma}}_{pn}\boldsymbol{v}_{n}=0\text{.}  \label{Vp0}
\end{equation}%
This equation tells us that the superfluid liquids move in such a way that the
effective velocity of the superfluid protons stay zero (insuring
conservation of the electric charge). In this case, it follows from Eq. (\ref%
{sing}) that the energy gap, in the proton spectrum, remains constant under
oscillations of the superfluid mixture of nucleons.

\subsection{Anisotropy of the $^3$P$_2$ order parameter in the superfluid flow}
\label{sec:an} 

Before proceeding to solutions for the mixture of superfluid  nucleons let us 
determine the most energetically favourable state of the triplet condensate in the neutron superfluid flow moving uniformly with the velocity 
$\tilde{V}_n$ at temperatures just below $T_{cn}$.

Assuming that the neutron gap is small, it is possible to expand Eq. (\ref%
{trip}) for the $^{3}$P$_{2}$ pairing in powers of $(\tilde{\Delta}^{\left(
n\right) }/T)^{2}\ll1$ and $(\mathbf{k\tilde{V}}_{n}/T)^{2}\ll1$ to get%
\begin{align}
\frac{7\zeta\left( 3\right) }{8\pi^{2}}\frac{\tilde{\Delta}^{\left( n\right)
2}}{T_{cn}^{2}}\left\langle \bar{b}^{4}(\mathbf{\hat{k})}\right\rangle & =%
\frac{T_{cn}-T}{T_{cn}}  \notag \\
& + \frac{7\zeta^{\prime}\left(-2\right)}{T_{cn}^{2}}\left\langle \bar {b}%
^{2}(\mathbf{\hat{k})}\,\left( \mathbf{k}_{Fn}\allowbreak\mathbf{\tilde {V}}%
_{n}\right) ^{2}\right\rangle ,  \label{v2}
\end{align}
where $\zeta^{\prime}$ is a derivative of the Riemann zeta function, $%
\zeta^{\prime}\left( -2\right) =-0.03045$. The angle brackets denote the
average over the Fermi surface,
\begin{equation}
\left\langle \cdot\cdot\cdot\right\rangle =\int\frac{d\mathbf{\hat{k}}}{4\pi 
}\cdot\cdot\cdot.  \label{ab}
\end{equation}

When the superfluid condensate is at rest relative the normal component the
preferred direction of the quantization axis can be chosen arbitrary. In
equilibrium at a non-zero temperature this leads to the formation of a loose
domain structure, where each domain has a preferred orientation, and these
domains are randomly oriented in space \citep{an58}. However, the superfluid
motion violates the degeneration over the directions of quantization axes in
different domains. To demonstrate this consider first a separate domain in
the moving superfluid of a size, much larger than the correlation length,
when the influence of the domain walls can be neglected \footnote{%
Near the domain walls, the unitary condition can be violated at distances of
the order of the correlation length \citep{r72}.}.

\begin{figure}
\includegraphics[width=\columnwidth]{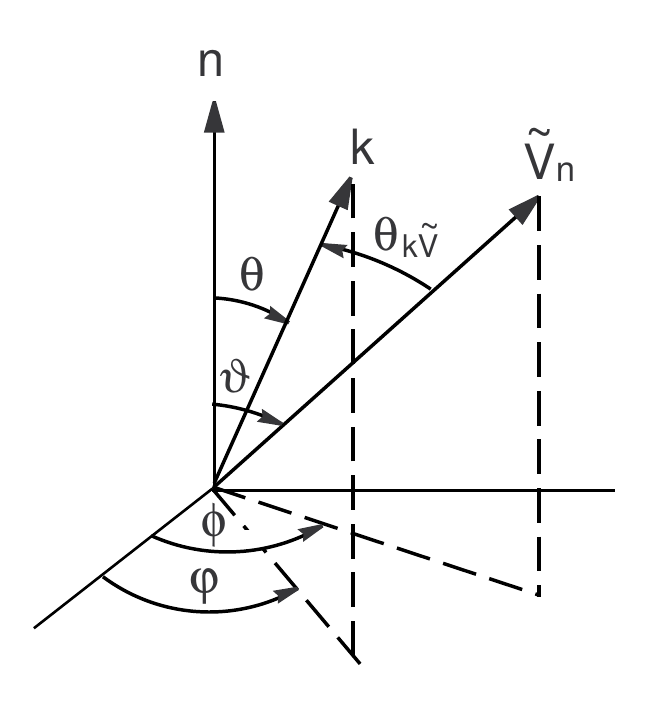}
\caption{A sketch of the geometry and the different coordinate systems that are used to address the minimum energy state in a moving triplet-paired neutron condensate.}
\label{fig:geom}
\end{figure}

Suppose the superfluid flow moves with the effective velocity $\mathbf{%
\tilde {V}}_{n}$ relative the normal component. In this case to perform the
angle averaging in Eq. (\ref{v2}) we introduce the spherical polar
coordinates $(k,\theta,\varphi)$ for the quasi-particle momentum $\mathbf{k}$
relative the preferred direction $\mathbf{n}$, and adopt that the vector $%
\mathbf{\tilde{V}}_{n}$ has spherical angle coordinates $\left( \vartheta,\phi\right) $
relative to the polar axis directed along $\mathbf{n}$, as depicted 
in Fig. \ref{fig:geom}. 
Than, by
the addition theorem, the angle $\theta_{\mathbf{k\tilde{V}}}$ between the
quasi-particle momentum and the flow direction can be written as 
\begin{equation}
\cos\theta_{\mathbf{k\tilde{V}}}=\cos\vartheta\cos\theta+
\sin\vartheta\sin\theta\cos\left( \varphi-\phi\right) ,  \label{et}
\end{equation}
and the average over directions of the quasi-particle momentum is given by
the integral%
\begin{equation*}
\left\langle \cdot\cdot\cdot\right\rangle =\frac{1}{4\pi}\int\sin
\theta\,d\theta\,d\varphi\cdot\cdot\cdot.
\end{equation*}

In what follows we abbreviate the quantum numbers $M=0$ and $\left\vert
M\right\vert =2$ as $M$. Taking into account the cylindrical symmetry of the
anisotropic energy gap about the quantization axis: $\bar{b}_M^{2}(\mathbf{%
\hat{k})}=\bar{b}_M^{2}\left( \cos\theta\right) $, from Eq. (\ref{v2}) we
obtain%
\begin{align}
\frac{\tilde{\Delta}_{M}^{\left( n\right) }}{T_{cn}} & =\left[ \frac{8\pi^{2}%
}{7\zeta\left( 3\right) }\frac{5}{6}\frac{T_{cn}-T}{T_{cn}}\right.  \notag \\
& \left. -\frac{5}{3}\frac{k_{Fn}^{2}}{T_{cn}^{2}}\allowbreak\tilde{V}%
_{n}^{2}\,\,\left( K_{1}^{\left( M\right) }\cos^{2}\vartheta +K_{2}^{\left(
M\right) }\sin^{2}\vartheta\right) \right] ^{1/2},  \label{dl}
\end{align}
where%
\begin{equation}
K_{1}^{\left( M\right) }=\int_{0}^{1}d\left( \cos\theta\right) \,\bar{b}%
_{M}^{2}\left( \cos\theta\right) \cos^{2}\theta\,  \label{K1}
\end{equation}
and%
\begin{equation}
K_{2}^{\left( M\right) }=\frac{1}{2}\int_{0}^{1}d\left( \cos \theta\right) \,%
\bar{b}_{M}^{2}\left( \cos\theta\right) \sin ^{2}\theta\,  \label{K2}
\end{equation}
When obtaining Eq. (\ref{dl}) the fact is used that $\left\langle \bar {b}%
_{M}^{4}(\mathbf{\hat{k})}\right\rangle =6/5$ in both the cases $M=0$ and $%
\left\vert M\right\vert =2$.

In order to choose the most favourable state of the moving
condensate, it is necessary to estimate the free energy of the superfluid
flow. The kinetic part of the free energy of the flow is the same for the 
$^{3}$P$_{2}$ states with $M=0$ and $\left\vert M\right\vert =2$. 
Therefore one has to
consider only the internal part of the free energy excess, that is, the
difference between the free energy of the superfluid state $F_{\mathtt{s}}$
and that of the normal state $F_{\mathtt{norm}}$. The corresponding expression for the case of spin-singlet pairing was obtained in  \citet{agd}. Following this work in the case of spin-triplet pairing for  $T-T_{cn}\ll T_{cn}$, we get
\begin{equation}
F_{s}-F_{\mathtt{norm}}=-\left( \frac{k_{F}m^{\ast }}{2\pi ^{2}}\right) 
\frac{7\zeta\left( 3\right) }{16\pi ^{2}}\frac{\tilde{\Delta}^{4}}{T^{2}}
\left\langle 
\bar{b}_{M}^{4}\right\rangle   \label{df}
\end{equation}%
This expression is very similar to that obtained for the case
of isotropic spin-singlet pairing, except of the factor $\left\langle \bar{b}%
_{M}^{4}\right\rangle $, which is caused by the anisotropy of the energy gap
in the $^{3}$P$_{2}$ state. As was mentioned above, this factor, $%
\left\langle \bar{b}_{M}^{4}\right\rangle =6/5$, is the same in both the
cases, $M=0$ and $\left\vert M\right\vert =2$. As can be seen, the excess
free energy is a monotonic function of the amplitude of the energy gap $%
\tilde{\Delta}$, whose dependence on the effective flow velocity is given in
Eq. (\ref{dl}). A larger energy gap corresponds to a lower energy of the moving condensate.

It follows from Eq. (\ref{dl}) that the total energy is minimal for $%
\vartheta=0$ or $\pi$ if $K_{1}^{\left( M\right) }<K_{2}^{\left( M\right) }$%
, and for $\vartheta=\pi/2$ if $K_{1}^{\left( M\right) }>K_{2}^{\left(
M\right) }$. For the pairing case with $\left\vert M\right\vert =2$ when $%
\bar{b}_{\left\vert M\right\vert =2}^{2}=(3/2)\sin ^{2}\theta$ the simple
integration yields%
\begin{equation}
K_{1}^{\left( \left\vert M\right\vert =2\right) }=\frac{1}{5},~\
K_{2}^{\left( \left\vert M\right\vert =2\right) }=\frac{2}{5}.  \label{K1n}
\end{equation}
For the pairing with $M=0$ when $\bar{b}_{M=0}^{2}=\left(1+3\cos^{2}\theta%
\right)/2$ one gets 
\begin{equation}
K_{1}^{\left( M=0\right) }=\frac{7}{15},~\ K_{2}^{\left( M=0\right) }=\frac{4%
}{15}.  \label{K2n}
\end{equation}

\begin{figure}
\includegraphics[width=\columnwidth]{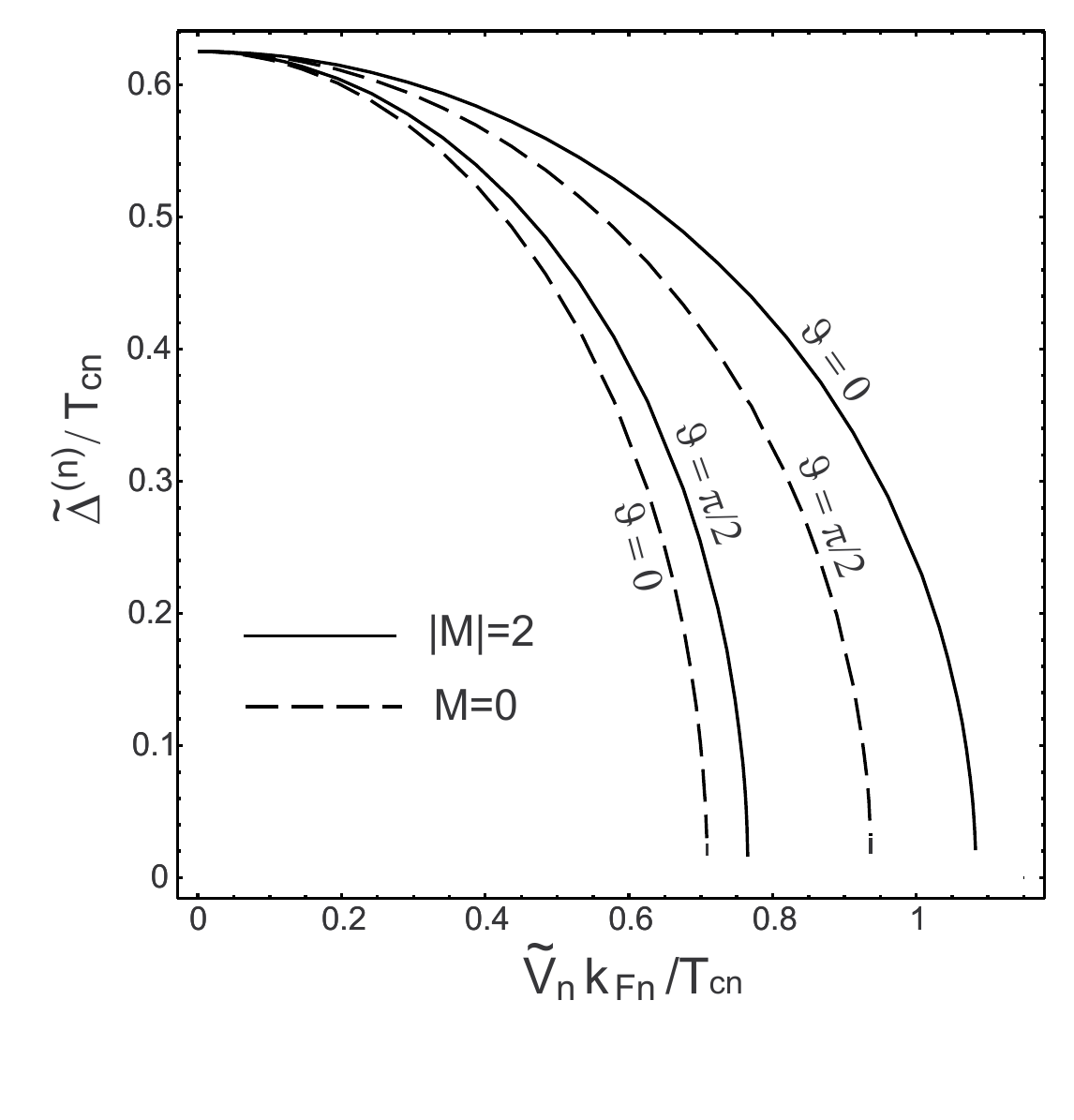}
\caption{The dependence of the energy gap amplitude on the effective
velocity $\tilde{V}_{n}$ for $\protect\vartheta=0$ and $\protect\vartheta=%
\protect\pi/2$. The solid lines are prepared for a spin-triplet condensate
with $\left\vert M\right\vert =2$, the dashed lines correspond to Cooper
pairs with $M=0$. All the curves are plotted for the dimensionless
temperature $\protect\tau=0.95$.}
\label{fig:figgap02}
\end{figure}

At fixed temperature the right-hand side of Eq. (\ref{dl}) depends on the
effective velocity $\tilde{V}_{n}$ of superfluid neutrons and is a monotonic
function of the angle $\vartheta$ between the quantization axis and the
direction of the effective flow velocity. Figure \ref{fig:figgap02} shows the dependence of
the energy gap amplitude on the effective velocity $\tilde{V}_{n}$ for $%
\vartheta=0$ and $\vartheta=\pi/2$. The solid lines are prepared for a
spin-triplet condensate with $\left\vert M\right\vert =2$, the dashed lines
correspond to Cooper pairs with $M=0$. All the curves are plotted for the
dimensionless temperature $\tau \equiv T/T_{cn}=0.95$.

As follows from this figure in an immovable superfluid liquid, all directions
of the principal axis of the gap matrix are equivalent, because the
corresponding states are degenerate. However, the superfluid motion
eliminates the degeneration over the directions of the quantization axis.
Moreover, states with $M=0$ and $\left\vert M\right\vert =2$, degenerate in
an immovable superfluid liquid, are also split by a superfluid motion. The
splitting of the energy levels rapidly increases with the increase in the
effective flow velocity.

From this figure one may conclude:

(1) At temperatures just below the critical temperature $T_{cn}$ the
superfluid motion makes the condensate with $\left\vert M\right\vert =2$
more energetically favourable than the condensate with $M=0$.

(2) The largest gap amplitude (the lowest flow energy) is realised when the
quantization axis is directed along the superfluid flow velocity.

Since at the fixed velocity the pairing always occurs into the state of
lowest energy, in what follows, we focus on the spin-triplet condensation
with $\left\vert M\right\vert =2$ assuming that the quantization axis is
directed along the flow direction. This justifies the general form (\ref%
{ksitpm}) of the correction to the quasi-particle energy caused by the
motion in superfluid neutron-proton mixture.

\begin{figure}
\includegraphics[width=\columnwidth]{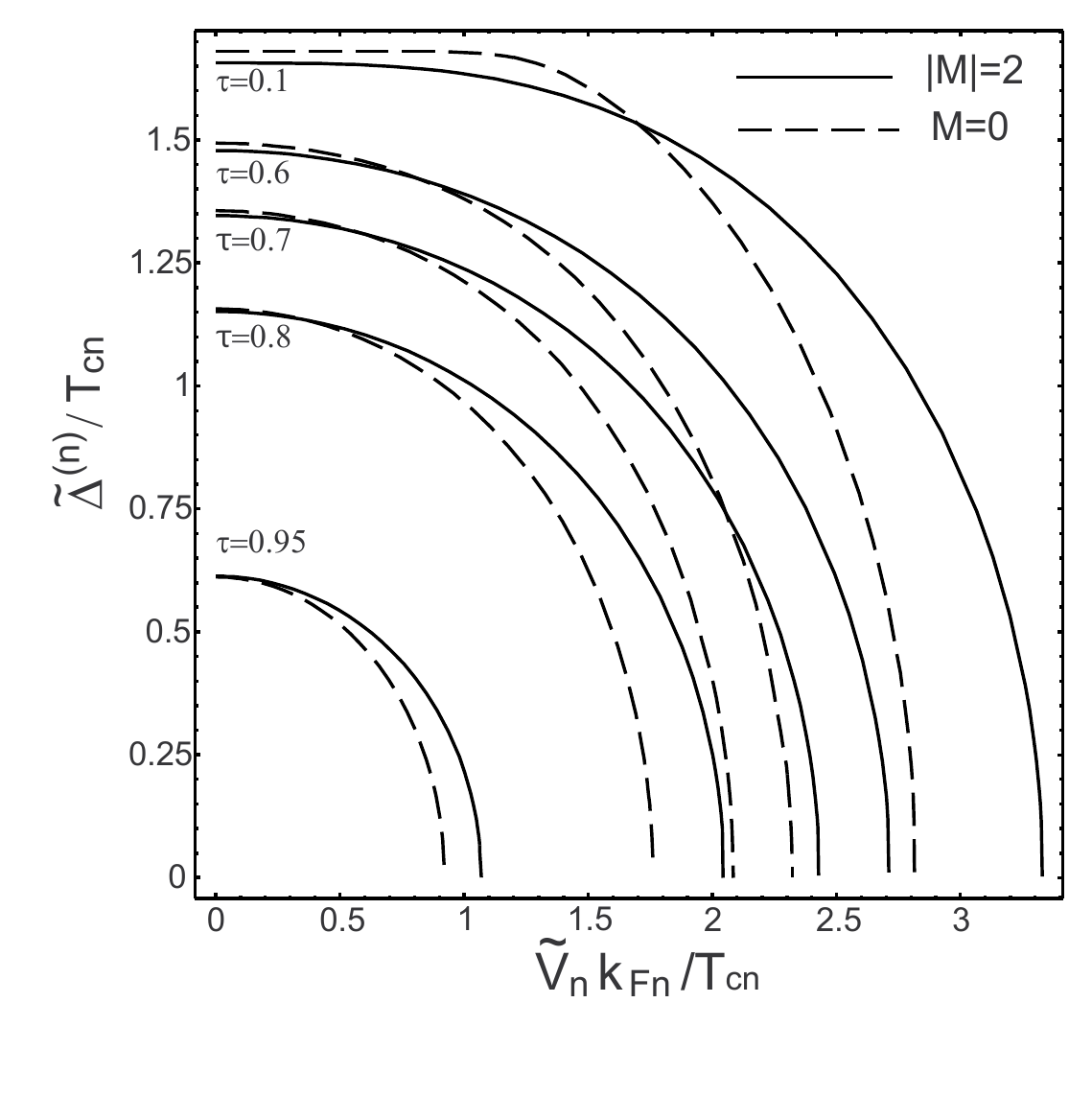}
\caption{The energy gap amplitude against the effective velocity
$\tilde{V}_{n}$
at different temperatures $\protect\tau $ for a moving condensate with $%
\left\vert M\right\vert =2$ and $\protect\vartheta =0$ (the quantization
axis along the flow velocity) and for the competing superfluid state with $%
M=0$ and $\protect\vartheta =\protect\pi /2$ (the quantization axis is
perpendicular the flow velocity),}
\label{fig:pl1}
\end{figure}

At lower temperatures such that $T_{cn}-T\sim T_{cn}$ the physical picture
is more complicated. As, at zero temperature, the ground state with $M=0$ is
preferable for a superfluid condensate at rest (see Appendix B) one can expect
that, at lower temperatures, there exists some domain of the superfluid
velocities and temperatures, where the state with $M=0$ has the lowest
energy. Fig. \ref{fig:pl1} shows the dependence of the energy gap amplitude
on the effective velocity $\tilde{V}_{n}$ at different temperatures $\tau $
for a moving condensate with $\left\vert M\right\vert =2$ and $\vartheta =0$
(the quantization axis along the flow velocity) and for the competitive
superfluid state with $M=0$ and $\vartheta =\pi /2$, when the quantization
axis is perpendicular the flow velocity. 

Numerical estimates show a strong competition of the $^{3}$P$_{2}$ states
with $M=0$ and $|M|=2$, which takes place near small superfluid velocities
below the temperature about $(0.6\div 0.7)T_{cn}$. Let us notice that, in
these domain of temperatures and superfluid velocities, the energy
difference of the competing states of the triplet condensate does not exceed
two percent, and the change of the gap value due to the superfluid motion
is relatively small. We therefore restrict our analysis to the neutron
condensation into the $^{3}$P$_{2}(|M|=2)$ state at arbitrary temperatures
below $T_{cn}$. We assume also that the quantization axis is directed along
the superfluid flow velocity, as it follows from the above arguing.

\section{A superfluid neutron-proton mixture with supercurrents.}

\label{sec:mix}

\subsection{Case of $(T_{cn}-T)/T_{cn}<<1$}

\label{sec:asn}

For temperatures just below the critical temperature for the neutron
superfluidity onset one can derive the analytic solution to the non-linear
equations for the entrainment matrix. At temperatures near $T_{cn}$, such
that $\left( \mathbf{k\tilde{V}}_{n}/T\right) ^{2}\ll 1$ and $\left( \tilde{%
\Delta}^{\left( n\right) }\left( T\right) /T\right) ^{2}\ll 1$, one can
expand in a series in these small parameters the distribution functions
under the integral in Eq. (\ref{Fi}) to obtain 
\begin{equation}
\tilde{\Phi}_{n}=3\int \frac{d\mathbf{\hat{k}}}{4\pi }\left( \mathbf{\hat{k}%
\hat{V}}_{n}\right) ^{2}\left( 1+7\zeta ^{\prime }\left( -2\right) \frac{%
\tilde{\Delta}^{\left( n\right) 2}}{T_{cn}^{2}}\bar{b}^{2}\left( \mathbf{%
\hat{k}}\right) \right) ,  \label{Fin}
\end{equation}%
where $\bar{b}^{2}=3/2\sin ^{2}\theta $ and $\mathbf{\hat{V}}_{n}$ is a
unit vector in the direction of the effective velocity of the neutron
superfluid flow, so that $\mathbf{\hat{k}}\mathbf{\hat{V}}_{n}=\cos
\theta $.

After performing the trivial integration over the solid angle we get%
\begin{equation}
\tilde{\Phi}_{n}=1+\frac{3}{5}7\zeta ^{\prime }\left( -2\right) \frac{\tilde{%
\Delta}^{\left( n\right) 2}\left( T\right) }{T_{cn}^{2}}.
\label{Find}
\end{equation}%

To obtain a consistent solution, we must substitute here the gap amplitude
from Eq. (\ref{dl}) with $\vartheta =0$ and with 
$K_{1}^{\left( \left\vert M\right\vert =2\right) }$ and 
$K_{2}^{\left( \left\vert M\right\vert =2\right) }$
from Eq (\ref{K1n}): 
\begin{equation}
\frac{\tilde{\Delta}_{|M|=2}^{\left( n\right) 2}}{T_{cn}^{2}}=\frac{8\pi
^{2}}{7\zeta \left( 3\right) }\frac{5}{6}\frac{T_{cn}-T}{T_{cn}}-\frac{1}{3}%
\frac{k_{Fn}^{2}}{T_{cn}^{2}}\allowbreak \tilde{V}_{n}^{2}.  \label{dl2}
\end{equation}%
Making use of the identity 
\begin{equation}
\zeta ^{\prime }\left( -2\right) \frac{8\pi ^{2}}{\zeta \left( 3\right) }=-2,
\label{id}
\end{equation}%
and combining Eqs. (\ref{Find}), (\ref{dl2}) and (\ref{dl}) we find 
\begin{equation}
\tilde{\Phi}_{n}=1-\left( 1-\tau \right) \left( 1-\frac{\tilde{V}_{n}^{2}}{%
V_{n\max }^{2}}\right) ,  \label{FIn}
\end{equation}%
where $\tau =T/T_{cn}$, and the maximal effective velocity is
\begin{equation}
V_{n\max }=\sqrt{\frac{20\pi ^{2}}{7\zeta \left( 3\right) }\left( 1-\tau
\right) }\frac{T_{cn}}{k_{Fn}}.  \label{Vnm}
\end{equation}%
Making use of Eq. (\ref{Tc}) we find 
\begin{equation}
T_{cn}=\Delta _{0}^{\left( n\right) }\frac{\sqrt{6}}{\pi }e^{C-5/6}\simeq
0.6034\Delta _{0}^{\left( n\right) }.  \label{tcn}
\end{equation}%
This allows us to write the maximal effective velocity in the form%
\begin{equation}
V_{n\max }=\sqrt{\frac{120e^{2C-5/3}}{7\zeta \left( 3\right) }\left( 1-\tau
\right) }\upsilon _{n}^{cr},  \label{tVnm}
\end{equation}%
where 
\begin{equation}
\upsilon _{n}^{cr}=\frac{\Delta _{0}^{\left( n\right) }}{k_{Fn}}.
\label{vncr}
\end{equation}

In Eq. (\ref{FIn}), the function $\tilde{\Phi}_{n}$ has written in terms of
the effective velocity of neutrons $\mathbf{\tilde{V}}_{n}$. To calculate the
entrainment matrix we need this function in terms of the superfluid
velocity of neutrons $\boldsymbol{v}_{n}$. To derive this function let us combine the equation 
\begin{equation}
\mathbf{\tilde{V}}_{n}=\mathsf{\tilde{\gamma}}_{nn}\boldsymbol{v}_{n}+\mathsf{%
\tilde{\gamma}}_{np}\boldsymbol{v}_{p},  \label{Vn}
\end{equation}%
which follows from Eq. (\ref{V}) with Eq. (\ref{Vp0}) to obtain 
\begin{equation}
\mathbf{\tilde{V}}_{n}=\frac{1}{\mathsf{\tilde{\gamma}}_{pp}}\left( \mathsf{%
\tilde{\gamma}}_{nn}\mathsf{\tilde{\gamma}}_{pp}-\mathsf{\tilde{\gamma}}_{np}%
\mathsf{\tilde{\gamma}}_{pn}\right) \boldsymbol{v}_{n}.  \label{Vofv}
\end{equation}%
Here the functions $\mathsf {\tilde{\gamma}}_{ab}$ are defined in the equations
(\ref {s1}) - (\ref {S}), where one has to put $\tilde {\Phi}_{p}=0$.

We now use the fact that the function $(1-\tilde{\Phi}_{n})$ is small as $%
\left( 1-\tau \right) \ll 1$ [see Eq. (\ref{FIn})]. Therefore, in Eq. (\ref%
{Vofv}) one can take $\tilde{\Phi}_{n}$ in the lowest (zero) order, by
replacing $\tilde{\Phi}_{n}\rightarrow 1$. From Eq. (\ref{tVnm}) we thus
obtain 
\begin{equation}
\frac{\tilde{V}_{n}}{V_{n\max }}=\sqrt{\frac{7\zeta \left( 3\right)
e^{5/3-2C}}{120\left( 1-\tau \right) }}\frac{m_{n}}{m_{n}^{\ast }}\frac{v_{n}%
}{\upsilon _{n}^{cr}}.  \label{Lamn}
\end{equation}%
Substituting this expression in Eq. (\ref{FIn}) we obtain the function  
\begin{equation}
\tilde{\Phi}_{n}\simeq \tau +\left( 1-\tau \right) \left( \frac{v_{n}}{%
v_{n}^{\max }}\right) ^{2},~v_{n}<v_{n}^{\max },  \label{Fina}
\end{equation}%
and $\tilde{\Phi}_{n}=1$ if $v_{n}\geq v_{n}^{\max }$, where 
\begin{equation}
v_{n}^{\max }=\frac{m_{n}^{\ast }}{m_{n}}\sqrt{\frac{120\left( 1-\tau
\right) }{7\zeta \left( 3\right) e^{5/3-2C}}}\upsilon _{n}^{cr}  \label{vmax}
\end{equation}%
is the velocity of the superfluid flow at which the neutron
superfluidity collapses at temperature $\tau $. 

Further calculation can be done with the aid of Eqs. (\ref{s1})-(\ref{S})
and (\ref{rab}), where one has to put $\tilde{\Phi}_{p}=0$, as the
superfluid protons are considered in the low-temperature limit. We thus
obtain 
\begin{equation}
{\rho }_{nn}=\rho _{n}\,\mathsf{\tilde{\gamma}}_{nn}\,\left( 1-\tilde{\Phi}%
_{n}\right) ,~\ {\rho }_{pp}=\rho _{p}\,\mathsf{\tilde{\gamma}}_{pp}\,,
\label{rnnpp}
\end{equation}%
and%
\begin{equation}
{\rho }_{pn}={\rho }_{np}=\rho _{n}\,\mathsf{\tilde{\gamma}}_{np}\,\left( 1-%
\tilde{\Phi}_{n}\right) ,  \label{rpnnp}
\end{equation}%
where

\begin{eqnarray}
\mathsf{\tilde{\gamma}}_{pp} &=&{\frac{m_{p}}{m_{p}^{\ast }}}\left( 1+{\frac{%
F_{1}^{pp}}{3}}-\frac{1}{9}\frac{F_{1}^{pn}F_{1}^{np}\tilde{\Phi}_{n}}{1+\,%
\tilde{\Phi}_{n}F_{1}^{nn}/3}\right)   \label{gpp1} \\
\mathsf{\tilde{\gamma}}_{nn} &=&{\frac{m_{n}}{m_{n}^{\ast }}\frac{%
1+F_{1}^{nn}/3}{1+\,\tilde{\Phi}_{n}F_{1}^{nn}/3}}  \label{gnn1} \\
\mathsf{\tilde{\gamma}}_{np} &=&{\frac{m_{p}}{\sqrt{m_{n}^{\ast
}\,m_{p}^{\ast }}}}\left( {\frac{k_{Fp}}{k_{Fn}}}\right) ^{3/2}{\frac{%
\,F_{1}^{np}/3}{1+\,\tilde{\Phi}_{n}F_{1}^{nn}/3}}\,.\,\,\,  \label{gnp1}
\end{eqnarray}

\subsection{General case}

\label{sec:gc}

Calculation of the non-linear entrainment matrix at arbitrary temperature of the mixture of superfluid nucleons to require numerical computations.
Let us write the function $\tilde{\Phi}_{n}$ and the gap
equation, defined in Eqs. (\ref{Fi}) and (\ref{trip}) as 
\begin{equation}
\tilde{\Phi}_{n}=\frac{3}{k_{Fn}\tilde{V}_{n}}\int \frac{d\mathbf{\hat{%
k}}}{4\pi }\int_{0}^{\infty }d\xi \left( \mathcal{\tilde{F}}_{-}^{\left(
n\right) }-\mathcal{\tilde{F}}_{+}^{\left( n\right) }\right) \cos \theta ,
\label{FinEq}
\end{equation}%
\begin{equation}
\ln \frac{\Delta _{0}^{\left( n\right) }}{\tilde{\Delta}^{\left( n\right) }}%
=\int \frac{d\mathbf{\hat{k}}}{4\pi }\int_{0}^{\infty }d\xi \frac{\bar{b}%
^{2}(\mathbf{\hat{k}})}{\tilde{E}_{\mathbf{k}}^{\left( n\right) }}\left( 
\mathcal{\tilde{F}}_{+}^{\left( n\right) }+\mathcal{\tilde{F}}_{-}^{\left(
n\right) }\right),   \label{DnEq}
\end{equation}%
where the distribution functions and the bogolon energy are of the form%
\begin{equation}
\mathcal{\tilde{F}}_{\pm }^{\left( n\right) }=\frac{1}{1+e^{\left( \tilde{E}%
_{\mathbf{k}}^{\left( n\right) }\pm k_{Fn}\tilde{V}_{n}\cos \theta \right)
/T}},  \label{Fn}
\end{equation}%
and%
\begin{equation}
\tilde{E}_{\mathbf{k}}^{\left( n\right) }=\sqrt{\xi ^{2}+\tilde{\Delta}%
^{\left( n\right) 2}\bar{b}^{2}},  \label{EEn}
\end{equation}%
respectively. Since the quantization axis $\mathbf{n}$ to be directed along the effective velocity $\mathbf{\tilde{V}}_n$, in Eq. (\ref{Fn}) it is assumed
$\theta \equiv \theta_{\mathbf{\hat{k}\tilde{V}}_n}= \theta_{\mathbf{\hat{k}n}}$.

Let us remind that the  $^3$P$_2$ states with $M=0$ and $|M|=2$, are very close to one other   below the temperature about $(0.6\div 0.7)T_{cn}$ when the superfluid velocity substantially smaller the critical value. The change of the gap value due to the superfluid motion in this domain is relatively small. Therefore it is sufficient to restrict our analysis to the neutron condensation into the  $^3$P$_2(|M|=2)$ state at arbitrary temperatures below $T_{cn}$.

\begin{figure}
\includegraphics[width=\columnwidth]{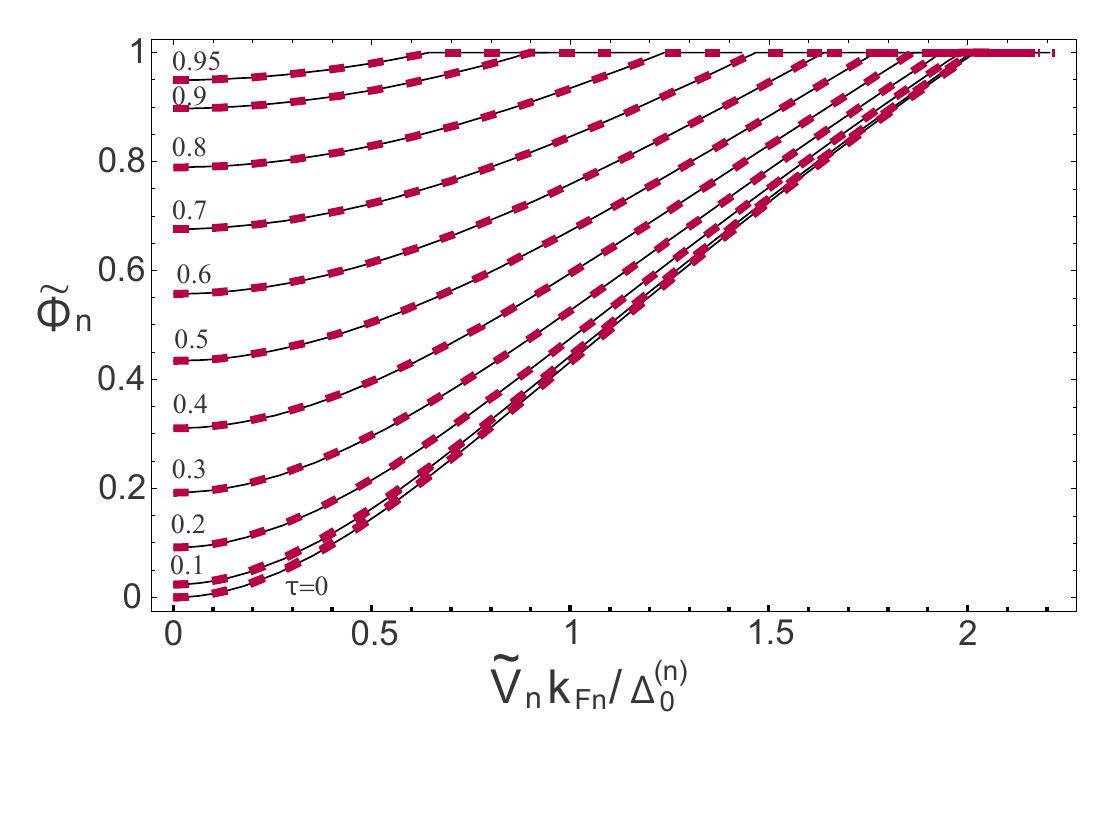}
\caption{The function $\tilde{\Phi}_{n}$ vs the dimensionless effective
velocity at different temperatures indicated near the curves in units $%
\protect\tau =T/T_{cn}$. Solid curves are as calculated numerically from a
self-consistent solution of Eqs. (\protect\ref{FinEq}) and 
(\protect\ref{DnEq}). The dashed lines are computed from the fit presented in 
Eqs.(\protect\ref{Fifit})-(\protect\ref{Vmax}).}
\label{fig:fin}
\end{figure}

The advantage of Eqs. (\ref{FinEq}) and (\ref{DnEq}) is that their solution 
$\tilde{\Phi}_{n}\left( T\right) $ is a function of the
temperature $T$ and absolute value of the 
effective velocity $\tilde{V}_{n}$, which (at this stage
of calculations) can be considered as external parameters. Self-consistent
numerical solutions to these equations are depicted in Fig. \ref{fig:fin} by
solid curves. The function $\tilde{\Phi}_{n}$ is plotted against $\tilde{V}%
_{n}$ for a set of dimensionless temperatures $\tau =T/T_{cn}$, ranging from 
$0$ to $1$.

The function $\tilde{\Phi}_{n}\left( T\right) $ is independent
explicitly of the Landau parameters. For practical calculations this
function can be fitted by the expressions:%
\begin{equation}
\tilde{\Phi}_{n}=f +\frac{g\,\left( 1-f\right) \left( \tilde{V}%
_{n}/V_{\max }\right) ^{2}}{1+\left( g-1\right) \left( \tilde{V}_{n}/V_{\max
}\right) ^{3/2}},~\ \tilde{V}_{n}<V_{\max }  \label{Fifit}
\end{equation}%
and%
\begin{equation}
\tilde{\Phi}_{n}=1,~\ \tilde{V}_{n}\geq V_{\max },  \label{Fifit1}
\end{equation}%
where%
\begin{equation}
f =\frac{2.3197\tau ^{2}\exp \left( 0.0082\sqrt{1-\tau }+0.4829\tau
\right) }{1+0.0788\tau +2.6808\tau ^{2}},  \label{fi}
\end{equation}%
\begin{equation}
g=\frac{2.9699+0.7847\tau +1.9295\tau ^{2}}{1+0.4844\tau +5.1598\tau ^{2}}.
\label{a}
\end{equation}%
The maximum effective velocity possible for the superfluid flow at the
temperature $\tau $ can be found from Eq. (\ref{DnEq}) in the limit $%
\tilde{\Delta}^{\left( n\right) }\rightarrow 0$. In units of $\Delta
_{0}^{(n)}/k_{Fn}$, it varies from $V_{\max }=\sqrt{3e/2}$ at $\tau =0$ to
zero at $\tau =1$, see \cite{Leinson2017}, and can be fitted as 
\begin{equation}
V_{\max }=2.0702\left( 1-\tau ^{2}\right) ^{1/2}-0.051\left( 1-\tau
^{2}\right) ^{2}.  \label{Vmax}
\end{equation}

Dashed lines in Fig. \ref{fig:fin} show the fitted function $\tilde{\Phi}%
_{n} $. As can be seen the fit describes the curves with a good accuracy, so
one can use it for a calculation of the entrainment matrix in superfluid
nucleon mixtures with an arbitrary set of the Landau parameters. We now
focus on this calculation.

\begin{figure}
\includegraphics[width=\columnwidth]{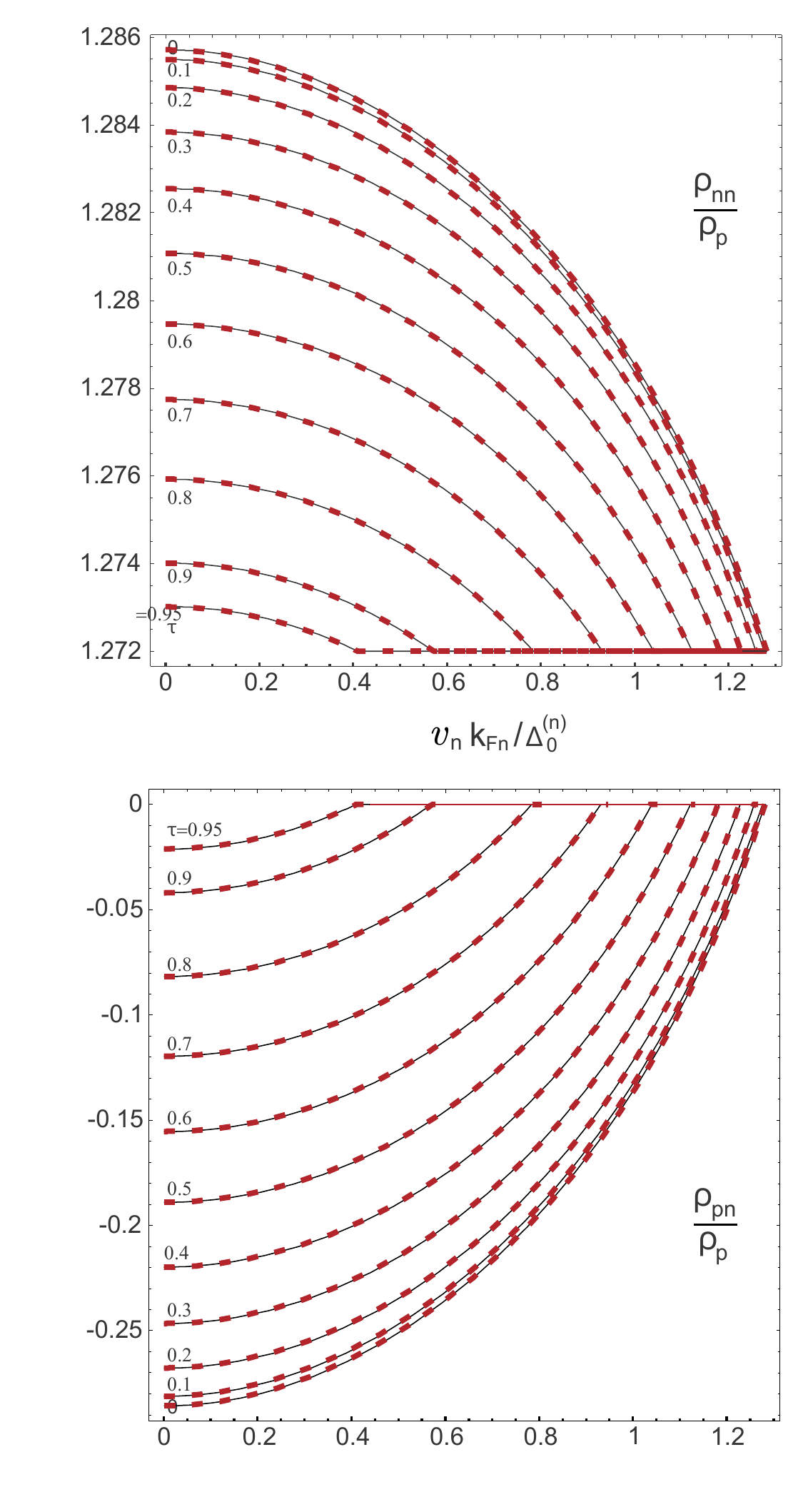}
\caption{The components of the entrainment matrix as functions of the neutron superfluid flow velocity $v_{n}$ in the mixture of $^{1}$S$_{0}$
superfluid protons and $^{3}$P$_{2}(|M|=2)$ superfluid neutrons for
different temperatures indicated near the curves in units $\protect\tau %
=T/T_{cn}$. The superfluid flow velocity is indicated in units of the
critical velocity $\Delta _{0}^{(n)}/k_{Fn}$. Solid curves demonstrate the
exact numerical result. The curves obtained with the aid of the fitted
expressions (\protect\ref{Fifit})-(\protect\ref{Vmax}) are shown in dashed
lines.}
\label{fig:ro}
\end{figure}
As we consider the proton superfluid in the low-temperature limit, the entrainment matrix is given by Eqs. (\ref{s1})-(\ref{S}) and (\ref{rab}), with 
$\tilde{\Phi}_{p}=0$. In this case, the right sides of these equations depend on the temperature and effective
velocity $\mathbf{\tilde{V}}_{n}$ of neutrons. We, however, need
the dependence of the entrainment matrix on the superfluid velocity 
$\boldsymbol{v}_{n}$.

To this end let us combine Eq. (\ref{Vn}), with Eq. (\ref{Vp0}) to obtain the 
superfluid velocity $\boldsymbol{v}_{n}$ as a function of the effective velocity 
$\mathbf{\tilde{V}}_{n}$. 
Making also use of Eqs. (\ref{s1})-(\ref{S}) and (\ref{rab}) 
with $\tilde{\Phi}_{p}=0$ we finally get
\begin{equation}
v_{n}={\frac{m_{n}^{\ast }}{m_{n}}}\frac{\left( 1+\,\tilde{\Phi}%
_{n}\,F_{1}^{nn}/3\right) \,\left( 1+F_{1}^{pp}/3\,\right) -\,\tilde{\Phi}%
_{n}\,F_{1}^{np}F_{1}^{pn}/9}{\left( 1+F_{1}^{pp}/3\right) \left(
1+F_{1}^{nn}/3\,\right) -F_{1}^{pn}F_{1}^{np}/9}\tilde{V}_{n},  \label{vofV}
\end{equation}%
\begin{eqnarray}
\frac{{\rho }_{nn}}{\rho _{n}} &=&{\frac{m_{n}}{m_{n}^{\ast }}\frac{%
1+F_{1}^{nn}/3}{1+\,\tilde{\Phi}_{n}F_{1}^{nn}/3}}(1-\tilde{\Phi}_{n})
\label{gnn} \\
\frac{{\rho }_{np}}{\rho _{n}} &=&{\frac{m_{p}}{\sqrt{m_{n}^{\ast
}\,m_{p}^{\ast }}}}\left( {\frac{k_{Fp}}{k_{Fn}}}\right) ^{3/2}{\frac{\left(
1-\tilde{\Phi}_{n}\right) \,F_{1}^{np}/3}{1+\,\tilde{\Phi}_{n}F_{1}^{nn}/3}}%
\,\,,\,\,  \label{gnp}
\end{eqnarray}%
and $\rho _{np}=\rho _{pn}$. Since we consider the proton superfluid in the
low-temperature limit the $\rho _{pp}$ matrix element can be obtained from
the identity due to Galilean invariance%
\begin{equation*}
\rho _{pp}+\rho _{pn}=\rho _{p}.
\end{equation*}

The right sides of Eqs. (\ref{vofV}), (\ref{gnn}) and (\ref{gnp}) depend on
the temperature and effective velocity $\tilde{V}_{n}$ which can take the
values between $0$ and $V_{\max }(\tau )$. The Eq. (\ref{vofV}) in a pair
with each of Eqs. (\ref{gnn}) and (\ref{gnp}) allows one to get the
parametric curves ${\rho }_{ab}(v_{n},T)$ employing $\tilde{V}_{n}$ as a
parameter.

For illustration, of the temperature and supercurrents effects on the
entrainment efficiency in the neutron star core we have considered a
superfluid nucleon liquid with a total density of baryons $%
n_{b}=n_{n}+n_{p}=2n_{0}$, where $n_{0}=0.16$ fm$^{-3}$ is the normal
nuclear density. We use realistic parameters obtained with the use of APR
equation of state \citep{apr98} in many calculations of a NS with the mass $%
1.4M_{\odot }$ \citep[see, e.g.][]{gkg14}. Assuming the
nucleon matter in beta-equilibrium the asymmetry parameter $\delta
=(n_{n}-n_{p})/n_{b}$ is adopted to be $\delta =0.7$. Following %
\citet{gkcg13} we employ the density-dependent model for spin-singlet
pairing of protons and spin-triplet pairing for neutrons in the NS core.
This model is in agreement with microscopic calculations 
\citep[see, e.g.][]{yls99} and is similar to the model of nucleon pairing used in
cooling simulations of the NS in Cassiopea A supernova remnant %
\citep{syhhp11}. 
For $n_{b}=2n_{0}$ this model suggests the critical temperatures for the
proton and neutron superfluidity onset $T_{cp}=6\times 10^{9}\,\mathsf{K}$
and $T_{cn}=5\times 10^{8}\,\mathsf{K}$, respectively. 

Information on the
Landau parameters for asymmetric nuclear matter is very limited. We use the
density-dependent Landau parameters obtained microscopically in %
\citet{gkh09a}, which turn out to be $F_{1}^{nn}=-1$, $F_{1}^{pp}=-0.3$, $%
F_{1}^{np}=F_{1}^{pn}=-0.24$ for $n_{b}=2n_{0}$.

The components of the entrainment matrix $\rho_{nn}$ and $\rho_{pn}$, as
functions of the superfluid velocity $v_{n}$ of the neutron flow, are
depicted in Fig. \ref{fig:ro}. Solid curves demonstrate the exact numerical
result. The curves obtained with the aid of the fitted expressions (\ref%
{Fifit})-(\ref{Vmax}) are shown in dashed lines.  
The $\rho_{pp}$ component of the entrainment matrix is almost independent of the
velocities of superfluid flows because the Eq. (\ref{Vp0}) insures that the
velocity of superfluid protons is small in comparison with its critical value%
\begin{equation*}
v_{p}\sim v_{n}<\Delta _{0}^{\left( n\right) }/k_{Fn}\ll \Delta_{0}^{\left(
p\right) }/k_{Fp}.
\end{equation*}%
This means that the the superfluid energy gap of protons remains constant at
this motion. In the considered case we got $\rho _{pp}=1.285\rho _p$.

Let us notice that the function $\tilde{\Phi}_{n}$, defined in Eq. (\ref%
{FinEq}) and fitted by the expressions (\ref{Fifit})-(\ref{Vmax})
universally depends on only the dimensionless temperature $\tau$ and
parameter $k_{Fn}\tilde{V}_{n}/\Delta _{0}^{\left( n\right) }$. This
function can be applied for any density of superfluid neutrons and any model
of the $^{3}$P$_{2}$ neutron pairing. Thus, a change of the entrainment
matrix $\rho _{ab}$, as the function of the temperature and parameter 
$k_{Fn}\tilde{V}_{n}/\Delta _{0}^{\left( n\right) }$, can be caused by only a
varying of the Landau parameters and the effective nucleon masses along
with the density change. To get an idea about the range of this changes let
us calculate the entrainment matrix for the nucleon density $n=3n_{0}$ in
the same model assumptions. In this case we take the asymmetry parameter in
beta-equilibrium to be $\delta =0.837$, the critical temperatures for the
proton and neutron superfluidity onset $T_{cp}=5\times 10^{9}\,\mathsf{K}$
and $T_{cn}=6\times 10^{8}\,\mathsf{K}$, respectively, and the Landau
parameters 
$F_{1}^{nn}=-1$, $F_{1}^{pp}=-0.55$, $F_{1}^{np}=F_{1}^{pn}=-0.3$.
\begin{figure}
\includegraphics[width=\columnwidth]{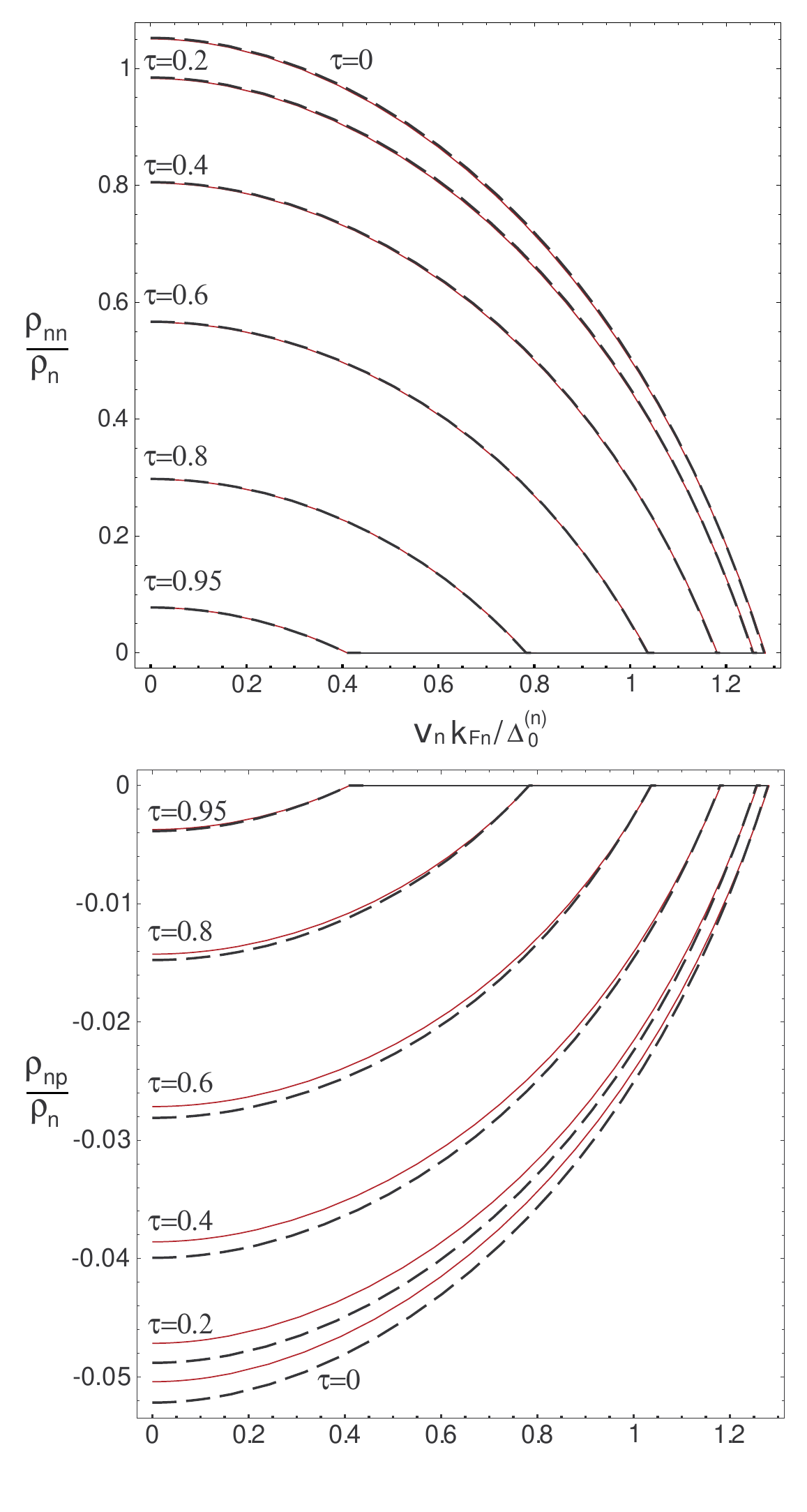}
\caption{The components of the entrainment matrix 
$\rho_{nn}$ and $\rho_{np}$, as
functions of the superfluid velocity $v_{n}$ of the neutron flow. Solid curves correspond to $n=2n_0$. 
The dashed lines are same but for the baryon density $n_b=3n_0$.}
\label{fig6}
\end{figure}
The result of this calculation is depicted in Fig. 6. 
For a comparison we show the curves calculated for $n_b=3n_0$ together with the results of the previous calculation for $n_b=2n_0$. One can see that the $\rho_{nn}$ component of the entrainment matrix is practically unchanged with the density increase, while the $\rho_{np}$ (and consequently $\rho_{pn}$) have slightly changed. Maximal change is about $4\%$

\section{Summary}

\label{sec:conc}

We considered the entrainment effect in a warm mixture of superfluid baryons in the core of an oscillating neutron star with an internal temperature below the critical temperature $T_{cn}$ for the onset of neutron superfluidity. It is assumed that the critical temperature for Cooper pairing  of protons  $T_{cp}$ is substantially higher than that for neutrons and thus protons are also superfluid. The suggested theory takes into account the entrainment dependence on the superfluid velocities and  temperature. The velocity dependence of the entrainment matrix is caused by a strong suppression of the superfluid energy gaps by supercurrents. Recently this effect was considered for the superfluid mixture of nucleons at zero temperature \citep{Leinson2017}. Calculations that also take into account temperature effects are proposed in the present work for the first time.

A closed system of non-linear equations for the entrainment matrix is considered taking into account the dependence of the superfluid gap on the temperature and velocities of superfluid nucleons.
Below the critical temperature $T_{cn}$, the entrainment matrix can be completely described by the Landau parameters and function $\tilde{\Phi}_{n}$ which universally depends on only the dimensionless temperature $\tau$ and parameter $k_{Fn}\tilde{V}_{n}/\Delta _{0}^{\left( n\right) }$. This function has been calculated numerically and fitted by simple formulas convenient for a practical use. It can be applied for any density of superfluid neutrons and any model of the $^{3}$P$_{2}$ neutron pairing.

The simple procedure is suggested for a construction of the entrainment matrix for arbitrary set of the Landau parameters $F_{1}^{nn}$, $F_{1}^{pp}$ and $F_{1}^{np}=F_{1}^{pn}$ with the aid of the fitted function $\tilde{\Phi}_{n}$ . The entrainment matrix, calculated in this way, is shown together with the exact numerical solution in Fig. \ref{fig:ro}, which demonstrates very good agreement.

From the plots it is seen also the strong dependence of the entrainment efficiency on the temperature and the neutron superfluid velocity. The value of superfluid velocity at which the neutron superfluidity is destroyed and the entrainment disappears crucially depends on the temperature and becomes very small at temperatures just below the critical value $T_{cn}$.  This effect significantly distinguishes our result from the entrainment matrix derived in \citet{gh05}, where the energy gaps and,  consequently, the entrainment matrix are assumed be independent of the superfluid velocities.

Let us notice that the simple expressions for a calculation of the entrainment matrix fitted in Eqs. (\ref{Fifit})-(\ref{Vmax}) of the present work are valid for any temperature below $T_{cn}$ and can be employed also for the case of very low temperature by substituting $\tau =0$. This limiting case is important for the most neutron stars older then  $\sim 10^5$ years. However, for oscillations of young neutron stars at the cooling epoch one has to take into consideration the temperature impact \citep[see, e.g.][]{ga06}.

The entrainment effect plays very important role in NS pulsations dynamics \citep{pr04,cch05,cc06}. Due to the complicated dependence on superfluid velocities, the entrainment matrix will become a non-linear function of the oscillation amplitude. Owing to the temperature effects, considered in the paper, the non-linearity will affect the pulsations in the warm superfluid mixture of nucleons at lower amplitudes than it takes place at zero temperature. The reduction of the neutron gap due to joint impact of the superfluid velocity and temperature should greatly influence the eigenfrequencies and eigenfunctions of oscillating NS. This will also influence the dissipation processes, because bulk viscosity of the nucleon superfluid mixture explicitly depends on  the entrainment matrix \citep{ars12}. The dependence of the entrainment matrix on temperature and on superfluid velocities is important for mutual friction and related phenomena. The potential possibility of these effects has already been noted in the work by \citet{gk13}.

In addition, one can expect that the destruction of neutron superfluidity caused by the critical supercurrents in oscillating neutron stars should alternate with the subsequent condensation of superfluid neutrons in the comoving reference frame (in our case in the rest frame). Indeed, as we have discussed, the destruction of the neutron superfluidity results in the bogolons forming the normal liquid, which should be  unstable with respect to Cooper pairing. As can be seen from Fig. \ref{fig:pl1}, below the critical temperature,  the neutron liquid is the most unstable to the condensation with the effective velocity $\mathbf{\tilde{V}}_{n}=0$, which corresponds to the superfluid velocity
$\boldsymbol{v}_{n}=-(\rho _{np}/\rho _{nn})\boldsymbol{v}_{p}$.
Under the influence of the pressure gradient and the force of gravity, the newly created superfluid liquid will be accelerated to the critical velocity and again destroyed.
This new regime of oscillations has not been discussed in the literature and deserves a separate study.


\appendix

\section{Single-particle Green function in a superfluid flow of fermions}

Consider a fermion system described by the following Hamiltonian with pairing
(For brevity, the volume of the system is set to $\Omega=1$.)
\begin{align}
\mathcal{H} &  =\sum_{\mathbf{k}\alpha}\xi_{\mathbf{k}}a_{\mathbf{k}\alpha
}^{\dagger}a_{\mathbf{k}\alpha}\nonumber\\
+ &  \frac{1}{2}\sum_{\substack{\mathbf{k}_{1}\alpha,\mathbf{k}_{2}%
\beta,\\\mathbf{k}_{3}\gamma,\mathbf{k}_{4}\delta}}\left\langle \mathbf{k}%
_{1}\alpha,\mathbf{k}_{2}\beta\left\vert U\right\vert \mathbf{k}_{4}%
\delta,\mathbf{k}_{3}\gamma\right\rangle a_{\mathbf{k}_{1}\alpha}^{\dagger
}a_{\mathbf{k}_{2}\beta}^{\dagger}a_{\mathbf{k}_{3}\gamma}a_{\mathbf{k}%
_{4}\delta}\label{H}%
\end{align}
where the Greek letters denote spin indices; $\xi_{\mathbf{k}}=p^{2}/2m-\mu$
and $\mu$ is the Fermi\textperiodcentered energy of the system. Since the
pairing interaction conserves the total momentum, its matrix element can be
written in the form
\begin{align}
&  \left\langle \mathbf{k}_{1}a,\mathbf{k}_{2}\beta\left\vert
U\right\vert \mathbf{k}_{4}\delta,\mathbf{k}_{3}\gamma\right\rangle
\nonumber\\
&  =\delta_{\mathbf{k}_{1}+\mathbf{k}_{2},\mathbf{k}_{3}+\mathbf{k}_{4}%
}U_{\alpha\gamma,\beta\delta}\left(  \frac{\mathbf{k}_{1}-\mathbf{k}_{2}}%
{2}\mathbf{,}\frac{\mathbf{k}_{4}-\mathbf{k}_{3}}{2}\right)  \label{U}%
\end{align}
Consider the case where the center of mass of all the pairs are moving with
velocity $\mathbf{v}_{s}$, that is, when there is a uniform flow of the
superfluid. If the total momentum of a Cooper pair at the point $\mathbf{R=}%
\left(  \mathbf{r}_{1}+\mathbf{r}_{2}\right)  /2$ is 
$2\mathbf{Q}=2m\mathbf{v}_{s}$,
the initial momenta of the pairing nucleons should be $\left(  \mathbf{k}%
+\mathbf{Q},-\mathbf{k}+\mathbf{Q}\right)  $ rather than with $\left(
\mathbf{k},-\mathbf{k}\right)  $. For this case the Gor'kov equations for the
ordinary and anomalous Green's functions have been derived in \cite{ft72}. In
particular, the equation for the ordinary Green function is of the following
form in the Matsubara representation
\begin{gather}
\left(  i\omega_{r}-\xi_{\mathbf{k}^{+}}\right)  \left(  i\omega_{r}%
+\xi_{\mathbf{k}^{-}}\right)  \mathcal{\tilde{G}}_{\alpha\beta}\left(
\omega_{r},\mathbf{k}\right)  -\sum_{\gamma\delta}\tilde{\Delta}_{\alpha
\gamma}\left(  \mathbf{\hat{k}}\right)  \nonumber\\
\times\tilde{\Delta}_{\gamma\delta}^{\dagger}\left(  \mathbf{\hat{k}}\right)
\mathcal{\tilde{G}}_{\delta\beta}\left(  \omega_{r},\mathbf{k}\right)
=\left(  i\omega_{r}+\xi_{\mathbf{k}^{-}}\right)  \delta_{\alpha\beta
},\label{GE}%
\end{gather}
Here and below $\mathbf{\hat{k}}=\mathbf{k}/k$,
\begin{equation}
\omega_{r}=\left(  2r+1\right)  \pi T,~\ r=1,2,3,\cdot\cdot\cdot\label{or}%
\end{equation}
is the fermion Matsubara frequency, and the tilde above a letter indicates
values that depend on the velocity of the superfluid flow. The order parameter
$\tilde{\Delta}_{\alpha\beta}$ represents a $2\times2$ matrix in spin space
$\left(  \alpha,\beta=\uparrow,\downarrow\right)  $, and
\begin{equation}
\xi_{\mathbf{k}^{\pm}}=\frac{\left(  \mathbf{k}\pm m\mathbf{v}_{s}\right)
^{2}}{2m}-\mu\simeq\xi_{k}\pm\mathbf{k}\mathbf{v}_{s}.\label{ksipm}%
\end{equation}%
In obtaining this equality we have neglected the term $mv_{s}^{2}/2$~assuming
$mv_{s}\ll k_{F}$, where the Fermi momentum is specified by the particle
number density $N$ of the degenerate Fermi gas%
\begin{equation}
k_{F}=\left(  3\pi^{2}N\right)  ^{1/3}.\label{kF}%
\end{equation}

If we restrict our consideration to the case of a non rotating neutron star
and consider the unitary states of the gap matrix \footnote{The unitary
condition implies that the superfluid state under consideration retains time
reversal symmetry and does not have, for example, spin polarization.} then%
\begin{equation}
\sum_{\gamma}\tilde{\Delta}_{\alpha\gamma}\left(  \mathbf{\hat{k}%
}\right)  \tilde{\Delta}_{\gamma\delta}^{\dagger}\left(
\mathbf{\hat{k}}\right)  =\delta_{\alpha,\delta}\tilde
{D}^{2\,}\left(  \mathbf{\hat{k}}\right)  ,\label{unit}%
\end{equation}
where $\tilde{D}^{2\,}\left(  \mathbf{\hat{k}}\right)  $ is real, and Eq.
(\ref{GE}) takes the simple form%
\begin{equation}
\mathcal{\tilde{G}}_{\alpha \beta }\left( \omega _{r},\mathbf{k}\right)
=\delta _{\alpha \beta }\frac{1}{2}\left( \frac{1+\xi _{k}/\tilde{E}_{%
\mathbf{k}}}{i\omega _{r}-\tilde{E}_{\mathbf{k}}-\mathbf{k}\mathbf{v}_{s}%
}+\frac{1-\xi _{k}/\tilde{E}_{\mathbf{k}}}{i\omega _{r}+\tilde{E}_{\mathbf{k}%
}-\mathbf{k}\mathbf{v}_{s}}\right) ,  \label{tgf}
\end{equation}%
where%
\begin{equation}
\tilde{E}_{\mathbf{k}}=\sqrt{\xi _{k}^{2}{}+\tilde{D}^{2\,}\left( \mathbf{%
\hat{k}}\right) }.  \label{etil}
\end{equation}%
The only change brought to these formulae by the supercurrent is that
instead of $i\omega _{r}$ we have $i\omega _{r}-\mathbf{k}\mathbf{v}_{s}$%
, and the chemical potential is shifted as well: $\mu \rightarrow \mu
-mv_{s}^{2}/2$.
\section{Spin-triplet pairing of neutrons}

As is well known the spin-triplet neutron condensate arises in the
high-density neutron matter mostly owing to the attractive spin-orbit and
tensor interactions in the channel of two quasi-particles \citet{Takatsuka}. 
However, the
tensor interactions are not very significant in the beta-stable baryon
matter [see, e.g. \citet{Takatsuka,Baldo92,Elg96,kcz01}]. Therefore to
avoid cumbersome calculations we restrict our analysis to the case of
neutron pairing in the unitary $^{3}$P$_{2}$ channel. In this case the
energy gap 
\begin{equation}
\tilde{D}_{\mathbf{\hat{k}}}=\tilde{\Delta}^{\left( n\right) }\bar{b}(%
\mathbf{\hat{k}})  
\label{D}
\end{equation}%
depends on the direction of the quasi-particle momentum $\mathbf{\hat{k}=k}%
/k $ and, in general, has nodes. The amplitude of the energy gap $\tilde{%
\Delta}^{\left( n\right) }$ must be real up to an arbitrary overall phase
factor. We, therefore, may adopt that the gap amplitude $\tilde{\Delta}%
^{\left( n\right) }$ is a real function which depends on the temperature $T$
and the effective flow velocity \textbf{$\tilde{V}$}$_{n}$. The gap
anisotropy is determined by some vector $\mathbf{\bar{b}}(\mathbf{\hat{k}})$
in the spin space, which is chosen to be real in accordance to the unitary
condition and is normalized by the condition 
\begin{equation}
\int \frac{d\mathbf{\hat{k}}}{4\pi }\bar{b}^{2}(\mathbf{\hat{k}})=1~.
\label{Norm}
\end{equation}%
It should be noted that, by virtue of Eq. (\ref{Norm}), the amplitudes $%
\Delta _{0}^{\left( n\right) }$ and $\tilde{\Delta}^{\left( n\right) }$ are
chosen as to represent the energy gap averaged over the solid angle. Defined
in this way, the average energy gap furnishes an overall measure of the
pairing correction to the ground-state energy in the preferred state.%
\footnote{%
It is necessary to notice that the definition of the gap amplitude is
ambiguous in the literature. For example, in the case of $\left\vert
M\right\vert =2$, our gap amplitude is $\sqrt{2/3}$ times larger than the
gap amplitude in \cite{YKL}. Ratio $\Delta _{0}^{\left( \left\vert
M\right\vert =2\right) }/T_{c}=1.6573$ differ in the same
proportion from those reported in \cite{ykgh01}.}

The equation that allows one to find the gap amplitude $\tilde{\Delta }%
^{\left( n\right) }$ as a function of the temperature and velocity \textbf{$%
\tilde{V}$}$_{n}$ of the superfluid flow can be written in the form 
\citet{ft72,Leinson2017}: 
\begin{equation}
\frac{k_{Fn}m_{n}^{\ast}}{\pi^{2}}\ln\frac{\Delta_{0}^{\left( n\right) }}{%
\tilde{\Delta}^{\left( n\right) }}=\sum_{\mathbf{k}}\frac{\bar{b}^{2}(%
\mathbf{\hat{k}})}{\tilde{E}_{\mathbf{k}}^{\left( n\right) }}\left( \mathcal{%
\tilde{F}}_{+}^{\left( n\right) }+\mathcal{\tilde{F}}_{-}^{\left( n\right)
}\right) .  \label{trip1}
\end{equation}
In this equation, $\Delta_{0}^{\left( n\right) }$ is the energy gap
amplitude in the superfluid at rest (i.e. for \textbf{$\tilde{V}$}$_{n}=0$)
and temperature $T=0$. The distribution functions for the Bogoliubov
excitations $\mathcal{\tilde{F}}_{\pm}^{\left( n\right) }$ are as defined in
Eq. (\ref{dis}) with%
\begin{equation}
\tilde{E}_{\mathbf{k}}^{\left( n\right) }=\sqrt{\xi_{k}^{\left( n\right) 2}+%
\tilde{\Delta}^{\left( n\right) 2}\bar{b}^{2}(\mathbf{\hat{k}})}.
\label{Ep1}
\end{equation}
Note that if one formally puts $\bar{b}\rightarrow1$, the equation (\ref%
{trip}) becomes identical to Eq. (\ref{sing}).

The vector $\mathbf{\bar{b}}(\mathbf{\hat{k}})$ defines the angle anisotropy
of energy gap which depends on the phase state of the superfluid condensate.
In general form, this vector can be written as $\bar{b}_{i}=\bar{A}_{ij}\hat{%
k}_{j}$, where $\bar{A}_{ij}$ is a $3\times3$ matrix. In the case of a
unitary $^{3}$P$_{2}$ condensate this matrix must be a real symmetric
traceless tensor. It may be specified by giving the orientation of its
principal axes and its two independent diagonal elements in its
principal-axis coordinate system.

In total, there are five matrices $\ A_{ij}^{\left( M\right) }$, in
accordance with the five possible projections $M=0,\pm1,\pm2$ of the total
angular momentum of a pair in the $^{3}$P$_{2}$ state. Since the ground
state should be invariant under time reversal, the states with magnetic
quantum numbers $\pm M$ must be populated with equal likelihood. This
requirement yields the following five symmetric combinations 
\begin{equation}
A^{\left( 0\right) }=\frac{1}{\sqrt{2}}\left( 
\begin{array}{ccc}
-1 & 0 & 0 \\ 
0 & -1 & 0 \\ 
0 & 0 & 2%
\end{array}
\right) ,  \label{A0}
\end{equation}%
\begin{equation}
A^{\left( 2\right) }+A^{\left( -2\right) }=\sqrt{\frac{3}{2}}\left( 
\begin{array}{ccc}
1 & 0 & 0 \\ 
0 & -1 & 0 \\ 
0 & 0 & 0%
\end{array}
\right) ,  \label{A2p}
\end{equation}%
\begin{equation}
A^{\left( 1\right) }-A^{\left( -1\right) }=-\sqrt{\frac{3}{2}}\left( 
\begin{array}{ccc}
0 & 0 & 1 \\ 
0 & 0 & 0 \\ 
1 & 0 & 0%
\end{array}
\right) ,~  \label{A1m}
\end{equation}%
\begin{equation}
A^{\left( 2\right) }-A^{\left( -2\right) }=i\sqrt{\frac{3}{2}}\left( 
\begin{array}{ccc}
0 & 1 & 0 \\ 
1 & 0 & 0 \\ 
0 & 0 & 0%
\end{array}
\right) ,~  \label{A2m}
\end{equation}%
\begin{equation}
\ A^{\left( 1\right) }+A^{\left( -1\right) }=-i\sqrt{\frac{3}{2}}\left( 
\begin{array}{ccc}
0 & 0 & 0 \\ 
0 & 0 & 1 \\ 
0 & 1 & 0%
\end{array}
\right) .  \label{A1p}
\end{equation}
On the other hand, the gap tensor $\bar{A}_{ij}$ should be diagonal, which
excludes the possibility of populating states with $M=\pm1$. \ Therefore
within the preferred coordinate system, there exist two simple solutions of
Eq. (\ref{trip1}) with $\bar{A}=A^{\left( 0\right) }$ and with $\bar {A}%
=A^{\left( 2\right) }+A^{\left( -2\right) }$. For the first solution, which
represents a condensation of the pairs into the state with $M=0$, we have $%
\bar{b}_{M=0}^{2}(\mathbf{\hat{k}})=1/2\left( 1+3\cos^{2}\theta\right) $.
The second solution corresponds to $\left\vert M\right\vert =2$. In this
case $\bar{b}_{\left\vert M\right\vert =2}^{2}(\mathbf{\hat{k}}%
)=3/2\sin^{2}\theta $. The $^{3}$P$_{2}$ condensates, with $M=0$ and $%
\left\vert M\right\vert =2$, are known to be almost degenerate in the
neutron superfluid at rest.

In Eq. (\ref{trip1}), $\Delta_{0}^{\left( a\right) }$ is uniquely related to
the temperature of the superfluid transition $T_{ca}$, which is assumed to
be known, and $\tilde{\Delta}^{\left( a\right) }$ is the energy gap, which
for a fixed density depends on the temperature and velocity \textbf{$\tilde{V%
}$}$_{a}$ of the superfluid flow relative the normal (non-superfluid )
component.

The critical temperature for the onset of superfluidity can be found from
Eq. (\ref{trip1}) with the flow velocity equal to zero, \textbf{$\tilde{V}$}$%
_{n}=0$. The summation on the right-hand side can be done with the aid of
the formula%
\begin{equation}
\int\frac{d^{3}k}{\left( 2\pi\right) ^{3}}\cdot\cdot\cdot=\frac{k_{F}m^{\ast}%
}{2\pi^{2}}\int\frac{d\mathbf{\hat{k}}}{4\pi}\int_{-\infty}^{\infty
}d\xi_{k}\cdot\cdot\cdot.  \label{1}
\end{equation}
Close to the transition point $T\rightarrow T_{cn}$, when $\Delta^{\left(
n\right) 2}/T^{2}\ll1$, the equation with \textbf{$\tilde{V}$}$_{n}=0$ takes
the form 
\begin{equation}
\ln\frac{\Delta_{0}^{\left( n\right) }}{T}=\ln\pi-C+\left\langle \bar{b}%
^{2}\ln\frac{1}{\bar{b}}\right\rangle +\frac{7\zeta\left( 3\right) }{8\pi^{2}%
}\frac{\Delta^{\left( n\right) 2}}{T^{2}},  \label{d0Tc}
\end{equation}
where $C=0.577216$ - Euler's constant, $\zeta\left( x\right) $ is the
Riemann zeta function, $\zeta\left( 3\right) =1.202$; and the angle brackets
denote the average over directions of the quasi-particle momentum, as
indicated in Eq. (\ref{ab}). Details of similar calculations can be found in %
\citet{lp80}. Additionally the fact is used that $\left\langle \bar{b}%
^{2}\right\rangle =1$, according to the normalization condition (\ref{Norm}%
). By Eq. (\ref{d0Tc}) the gap vanishes at the temperature%
\begin{equation}
T_{cn}=\Delta_{0}^{\left( n\right) }\frac{e^{C}}{\pi}\exp\left\langle \bar{b}%
^{2}\ln\bar{b}\right\rangle .  \label{Tc}
\end{equation}
In the case of $^{3}$P$_{2}$ pairing with $M=0$ this equation gives $%
\Delta_{0}^{\left( n\right) }=1.\,\allowbreak681T_{cn}~$while for $%
\left\vert M\right\vert =2$~one gets $\Delta_{0}^{\left( n\right)
}=1.\noindent657T_{cn}.$

Note that if one formally puts $\bar{b}$ equal to unity, the equation (\ref%
{Tc}) recover the well-known results of the $^{1}$S$_{0}$ pairing, presented
in Eq. (\ref{dp0}).
\begin{figure}
\includegraphics[width=\columnwidth]{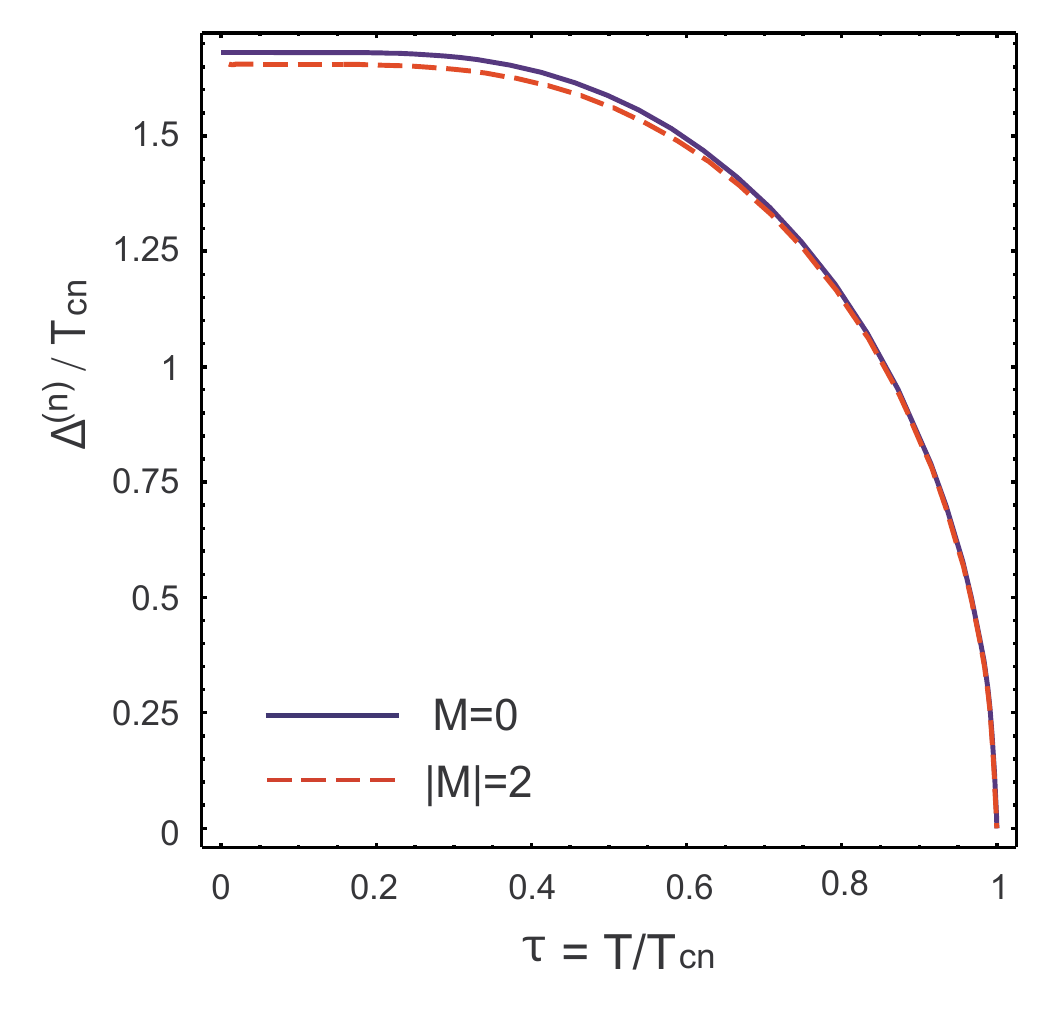}
\caption{The temperature dependence of the energy gap in the superfluid at
rest found from Eq. (\protect\ref{trip1}) with $\boldsymbol{v}_{n}=0$. The gap
amplitudes $\Delta^{\left( n\right) }$ for a spin-triplet condensation with $%
M=0$ and $\left\vert M\right\vert =2$ are shown in units of the critical
temperature $T_{cn}$ against the dimensionless temperature 
$\protect\tau =T/T_{cn}$.}
\label{fig:figD02}
\end{figure}

The temperature dependence of the energy gap in the superfluid at rest can
be found from Eq. (\ref{trip1}) with \textbf{$\tilde{V}$}$_{n}=0$. Numerical
solution to this equation is depicted in Fig. \ref{fig:figD02}, where the
energy gap amplitude $\Delta ^{\left( n\right) }$ is shown in units of the
critical temperature $T_{cn}$ versus the dimensionless temperature $\tau
=T/T_{cn}$ for a spin-triplet condensation with $M=0$ and $\left\vert
M\right\vert =2$. It can be seen that in an immovable superfluid liquid, the
two condensate states have a very close energies. We can consider them
degenerate in the temperature range just below the critical value.


\begin{thebibliography}{}
\makeatletter
\relax
\def\mn@urlcharsother{\let\do\@makeother \do\$\do\&\do\#\do\^\do\_\do\%\do\~}
\def\mn@doi{\begingroup\mn@urlcharsother \@ifnextchar [ {\mn@doi@}
  {\mn@doi@[]}}
\def\mn@doi@[#1]#2{\def\@tempa{#1}\ifx\@tempa\@empty \href
  {http://dx.doi.org/#2} {doi:#2}\else \href {http://dx.doi.org/#2} {#1}\fi
  \endgroup}
\def\mn@eprint#1#2{\mn@eprint@#1:#2::\@nil}
\def\mn@eprint@arXiv#1{\href {http://arxiv.org/abs/#1} {{\tt arXiv:#1}}}
\def\mn@eprint@dblp#1{\href {http://dblp.uni-trier.de/rec/bibtex/#1.xml}
  {dblp:#1}}
\def\mn@eprint@#1:#2:#3:#4\@nil{\def\@tempa {#1}\def\@tempb {#2}\def\@tempc
  {#3}\ifx \@tempc \@empty \let \@tempc \@tempb \let \@tempb \@tempa \fi \ifx
  \@tempb \@empty \def\@tempb {arXiv}\fi \@ifundefined
  {mn@eprint@\@tempb}{\@tempb:\@tempc}{\expandafter \expandafter \csname
  mn@eprint@\@tempb\endcsname \expandafter{\@tempc}}}

\bibitem[\protect\citeauthoryear{Abrikosov, Gor'kov  \&
  Dzyaloshinski}{Abrikosov et~al.}{1965}]{agd}
Abrikosov A.~A.,  Gor'kov L.~P.,   Dzyaloshinski I.~E.,  1965, Methods of
  Quantum Field Theory in Statistical Physics, 2nd ed..
Pergamon Press

\bibitem[\protect\citeauthoryear{Akmal, Pandharipande  \& Ravenhall}{Akmal
  et~al.}{1998}]{apr98}
Akmal A.,  Pandharipande V.~R.,   Ravenhall D.~G.,  1998, Phys. Rev. C58, 1804

\bibitem[\protect\citeauthoryear{Alexandrov}{Alexandrov}{2003}]{alex03}
Alexandrov A.~S.,  2003, Theory of Superconductivity: From Weak to Strong
  Coupling.
IOP publishing. Bristol \& Philadelphia

\bibitem[\protect\citeauthoryear{Alford, Reddy  \& Schwenzer}{Alford
  et~al.}{2012}]{ars12}
Alford M.~G.,  Reddy S.,   Schwenzer K.,  2012, Phys. Rev. Lett. 108, 111102

\bibitem[\protect\citeauthoryear{Alpar, Langer  \& Sauls}{Alpar
  et~al.}{1984}]{als84}
Alpar M.~A.,  Langer S.~A.,   Sauls J.~A.,  1984, Astrophys. J. 282, 533

\bibitem[\protect\citeauthoryear{Andersson}{Andersson}{1958}]{an58}
Andersson P.~V.,  1958, Phys. Rev. 112, 1900

\bibitem[\protect\citeauthoryear{Andersson}{Andersson}{1998}]{a98}
Andersson N.,  1998, ApJ, 502, 708

\bibitem[\protect\citeauthoryear{Andersson \& Itoh}{Andersson \&
  Itoh}{1975}]{ai75}
Andersson P.~V.,  Itoh N.,  1975, Nature, 256, 25

\bibitem[\protect\citeauthoryear{Andersson \& Kokkotas}{Andersson \&
  Kokkotas}{2001}]{ak01}
Andersson N.,  Kokkotas K.~D.,  2001, Int. J. of Modern Phys. D10, 381

\bibitem[\protect\citeauthoryear{Andersson, Kokkotas  \& Stergioulas}{Andersson
  et~al.}{1999}]{aks98}
Andersson N.,  Kokkotas K.~D.,   Stergioulas N.,  1999, Astrophys. J. 516, 307

\bibitem[\protect\citeauthoryear{Andersson, Comer  \& Prix}{Andersson
  et~al.}{2003}]{acp02}
Andersson N.,  Comer G.~L.,   Prix R.,  2003, Phys. Rev. Lett. 90, 091101

\bibitem[\protect\citeauthoryear{Andreev \& Bashkin}{Andreev \&
  Bashkin}{1975}]{ab75}
Andreev A.~F.,  Bashkin E.~P.,  1975, Zh. Exp. Teor. Fiz. 69, 319

\bibitem[\protect\citeauthoryear{Arras, Flanagan, Morsink, Schenk, Teukolsky
  \& Wasserman}{Arras et~al.}{2003}]{aetal03}
Arras P.,  Flanagan E.~E.,  Morsink S.~M.,  Schenk A.~K.,  Teukolsky S.~A.,
  Wasserman I.,  2003, ApJ, 591, 1129

\bibitem[\protect\citeauthoryear{Baldo, Cugnon, Lejeune  \& Lombardo}{Baldo
  et~al.}{1992}]{Baldo92}
Baldo M.,  Cugnon J.,  Lejeune A.,   Lombardo U.,  1992, Nucl. Phys. A536
  and349

\bibitem[\protect\citeauthoryear{Bardeen}{Bardeen}{1962}]{bardeen62}
Bardeen J.,  1962, Rev. of Modern Phys. 34, 667

\bibitem[\protect\citeauthoryear{Borumand, Joynt  \& Klu{\'{z}}niak}{Borumand
  et~al.}{1996}]{Betal96}
Borumand M.,  Joynt R.,   Klu{\'{z}}niak W.,  1996, Phys. Rev. C, 54, 2745

\bibitem[\protect\citeauthoryear{Brown, Cumming, Fattoyev, Horowitz, Page  \&
  Reddy}{Brown et~al.}{2017}]{bcf17}
Brown E.~F.,  Cumming A.,  Fattoyev F.~J.,  Horowitz C.~J.,  Page D.,   Reddy
  S.,  2017, Phys. Rev. Lett. 120 , 182701

\bibitem[\protect\citeauthoryear{Caillon, Gabinski  \& Labarsouque}{Caillon
  et~al.}{2001}]{cgl01}
Caillon J.~C.,  Gabinski P.,   Labarsouque J.,  2001, Nucl. Phys. A, 696, 623

\bibitem[\protect\citeauthoryear{Caillon, Gabinski  \& Labarsouque}{Caillon
  et~al.}{2002}]{cgl02}
Caillon J.~C.,  Gabinski P.,   Labarsouque J.,  2002, J. Phys. G, 28, 189

\bibitem[\protect\citeauthoryear{Caillon, Gabinski  \& Labarsouque}{Caillon
  et~al.}{2003}]{cgl03}
Caillon J.~C.,  Gabinski P.,   Labarsouque J.,  2003, J. Phys. G, 29, 2291

\bibitem[\protect\citeauthoryear{Carter, Chamel  \& Haensel}{Carter
  et~al.}{2005}]{cch05}
Carter B.,  Chamel N.,   Haensel P.,  2005, Nucl. Phys. A, 748, 675

\bibitem[\protect\citeauthoryear{Chamel \& Carter}{Chamel \&
  Carter}{2006}]{cc06}
Chamel N.,  Carter B.,  2006, Mon. Not. Roy. Astron. Soc. 368, 796

\bibitem[\protect\citeauthoryear{Chamel \& Haensel}{Chamel \&
  Haensel}{2006}]{ch06}
Chamel N.,  Haensel P.,  2006, Phys. Rev. C, 73, 045802

\bibitem[\protect\citeauthoryear{Chen, Clark, Dav\'{e}  \& Khodel}{Chen
  et~al.}{1991}]{ccdk91}
Chen J.,  Clark J.,  Dav\'{e} R.,   Khodel V.,  1991, Nucl. Phys. A555, 59

\bibitem[\protect\citeauthoryear{Chugunov \& Gusakov}{Chugunov \&
  Gusakov}{2011}]{chg11}
Chugunov A.~I.,  Gusakov M.~E.,  2011, Mon. Not. R. Astron. Soc. 418, L54

\bibitem[\protect\citeauthoryear{Comer \& Joynt}{Comer \& Joynt}{2003}]{cj03}
Comer G.~L.,  Joynt R.,  2003, Phys. Rev. D, 68, 023002

\bibitem[\protect\citeauthoryear{Elgaroy, Engvik, Hjorth-Jensen  \&
  Osnes}{Elgaroy et~al.}{1996a}]{eeho96}
Elgaroy O.,  Engvik L.,  Hjorth-Jensen M.,   Osnes E.,  1996a, Nucl. Phys.
  A604, 466

\bibitem[\protect\citeauthoryear{Elgaroy, Engvik, Hjorth-Jensen  \&
  Osnes}{Elgaroy et~al.}{1996b}]{Elg96}
Elgaroy O.,  Engvik L.,  Hjorth-Jensen M.,   Osnes E.,  1996b, Phys. Rev. Lett.
  77, 1428

\bibitem[\protect\citeauthoryear{Friedman \& Morsink}{Friedman \&
  Morsink}{1998}]{fm98}
Friedman J.~L.,  Morsink S.~M.,  1998, ApJ, 502, 714

\bibitem[\protect\citeauthoryear{Fujita \& Tsuneto}{Fujita \&
  Tsuneto}{1972}]{ft72}
Fujita T.,  Tsuneto T.,  1972, Prog. Theor. Phys. 48, 766

\bibitem[\protect\citeauthoryear{Gualtieri, Kantor  \& Gusakov}{Gualtieri
  et~al.}{2014}]{gkg14}
Gualtieri L.,  Kantor E.~M.,   Gusakov M.~E.,  2014, Phys.Rev. D90, 024010

\bibitem[\protect\citeauthoryear{Gusakov \& Andersson}{Gusakov \&
  Andersson}{2006}]{ga06}
Gusakov M.~E.,  Andersson N.,  2006, Mon. Not. R. Astron. Soc. 372, 1776

\bibitem[\protect\citeauthoryear{Gusakov \& Haensel}{Gusakov \&
  Haensel}{2005}]{gh05}
Gusakov M.~E.,  Haensel P.,  2005, Nucl.Phys. A, 761, 333

\bibitem[\protect\citeauthoryear{Gusakov \& Kantor}{Gusakov \&
  Kantor}{2013}]{gk13}
Gusakov M.~E.,  Kantor E.~M.,  2013, MNRAS, 428, L26

\bibitem[\protect\citeauthoryear{Gusakov, Yakovlev  \& Gnedin}{Gusakov
  et~al.}{2005}]{gyg05}
Gusakov M.~E.,  Yakovlev D.~G.,   Gnedin O.~Y.,  2005, MNRAS, 361, 1415

\bibitem[\protect\citeauthoryear{Gusakov, Kantor  \& Haensel}{Gusakov
  et~al.}{2009a}]{gkh09}
Gusakov M.~E.,  Kantor E.~M.,   Haensel P.,  2009a, Phys. Rev. C, 79, 055806

\bibitem[\protect\citeauthoryear{Gusakov, Kantor  \& Haensel}{Gusakov
  et~al.}{2009b}]{gkh09a}
Gusakov M.~E.,  Kantor E.~M.,   Haensel P.,  2009b, Phys. Rev. C, 80, 015803

\bibitem[\protect\citeauthoryear{Gusakov, Kantor, Chugunov  \&
  Gualtieri}{Gusakov et~al.}{2013}]{gkcg13}
Gusakov M.~E.,  Kantor E.~M.,  Chugunov A.~I.,   Gualtieri L.,  2013, MNRAS,
  428, 1518

\bibitem[\protect\citeauthoryear{Gusakov, Chugunov  \& Kantor}{Gusakov
  et~al.}{2014}]{gck14}
Gusakov M.~E.,  Chugunov A.~I.,   Kantor E.~M.,  2014, Phys. Rev. Lett. 112,
  151101

\bibitem[\protect\citeauthoryear{Haensel, Levenfish  \& Yakovlev}{Haensel
  et~al.}{2000}]{hly00}
Haensel P.,  Levenfish K.~P.,   Yakovlev D.~G.,  2000, A\&A, 357, 1157

\bibitem[\protect\citeauthoryear{Haensel, Levenfish  \& Yakovlev}{Haensel
  et~al.}{2001}]{hly01}
Haensel P.,  Levenfish K.~P.,   Yakovlev D.~G.,  2001, A\&A, 372, 130

\bibitem[\protect\citeauthoryear{Henning \& Manakos}{Henning \&
  Manakos}{1987}]{hm87}
Henning P.~A.,  Manakos P.,  1987, Nucl. Phys. A, 466, 487

\bibitem[\protect\citeauthoryear{Khodel, Clark  \& Zverev}{Khodel
  et~al.}{2001}]{kcz01}
Khodel V.~A.,  Clark J.~W.,   Zverev M.,  2001, Phys. Rev. Lett. 87, 031103

\bibitem[\protect\citeauthoryear{Leggett}{Leggett}{1965}]{leg65}
Leggett A.~J.,  1965, Phys. Rev. A, 140, 1869

\bibitem[\protect\citeauthoryear{Leggett}{Leggett}{1975}]{leg75}
Leggett A.~J.,  1975, Rev. of Modern Phys. 47, 331

\bibitem[\protect\citeauthoryear{Leinson}{Leinson}{2017}]{Leinson2017}
Leinson L.~B.,  2017, Mon. Not. Roy. Astron. Soc. 470. 3374

\bibitem[\protect\citeauthoryear{Lifshitz \& Pitaevskii}{Lifshitz \&
  Pitaevskii}{1980}]{lp80}
Lifshitz E.~M.,  Pitaevskii L.~P.,  1980, Statistical Physics, Part 2.
Pergamon, Oxford

\bibitem[\protect\citeauthoryear{Matsui}{Matsui}{1981}]{matsui81}
Matsui T.,  1981, Nucl. Phys. A, 370, 365

\bibitem[\protect\citeauthoryear{Mendell}{Mendell}{1991}]{m91}
Mendell G.,  1991, ApJ, 380, 515

\bibitem[\protect\citeauthoryear{Prix}{Prix}{2004}]{pr04}
Prix R.,  2004, Phys. Rev. D, 69, 043001

\bibitem[\protect\citeauthoryear{Prix \& Rieutord}{Prix \&
  Rieutord}{2002}]{pr02}
Prix R.,  Rieutord M. L.~E.,  2002, A\&A, 393, 949

\bibitem[\protect\citeauthoryear{Prix, Comer  \& Andersson}{Prix
  et~al.}{2002}]{pca02}
Prix R.,  Comer G.~L.,   Andersson N.,  2002, A\&A, 381, 178

\bibitem[\protect\citeauthoryear{Richardson}{Richardson}{1972}]{r72}
Richardson R.~W.,  1972, Phys. Rev. D, 5, 1883

\bibitem[\protect\citeauthoryear{Shternin \& Yakovlev}{Shternin \&
  Yakovlev}{2008}]{sy08}
Shternin P.~S.,  Yakovlev D.~G.,  2008, Phys. Rev. D78, 063006

\bibitem[\protect\citeauthoryear{Shternin, Yakovlev, Heinke, Ho  \&
  Patnaude}{Shternin et~al.}{2011}]{syhhp11}
Shternin P.~S.,  Yakovlev D.~G.,  Heinke C.~O.,  Ho W. C.~G.,   Patnaude D.~J.,
   2011, Mon. Not. Roy. Astron. Soc. 412, L108

\bibitem[\protect\citeauthoryear{Sidery, Passamonti  \& Andersson}{Sidery
  et~al.}{2010}]{spa10}
Sidery T.,  Passamonti A.,   Andersson N.,  2010, MNRAS, 405, 1061

\bibitem[\protect\citeauthoryear{Sj{\"{o}}berg}{Sj{\"{o}}berg}{1973}]{S73}
Sj{\"{o}}berg O.,  1973, Ann. Phys. 78, 39

\bibitem[\protect\citeauthoryear{Takatsuka}{Takatsuka}{1972}]{Takatsuka}
Takatsuka T.,  1972, Prog. Theor. Phys. 48, 1517

\bibitem[\protect\citeauthoryear{Takatsuka \& Tamagaki}{Takatsuka \&
  Tamagaki}{2004}]{tt04}
Takatsuka T.,  Tamagaki R.,  2004, Prog. Theor. Phys. 112, 37

\bibitem[\protect\citeauthoryear{Takatsuka, Nishizaki, Yamamoto  \&
  Tamagaki}{Takatsuka et~al.}{2001}]{tnyt01}
Takatsuka T.,  Nishizaki S.,  Yamamoto Y.,   Tamagaki R.,  2001, Nuclear
  Physics A691, 254c

\bibitem[\protect\citeauthoryear{Tamagaki}{Tamagaki}{1970}]{Tamagaki}
Tamagaki R.,  1970, Prog. Theor. Phys. 44, 905

\bibitem[\protect\citeauthoryear{Wang \& Shen}{Wang \& Shen}{2010}]{ws10}
Wang Y.~N.,  Shen H.,  2010, Phys. Rev. C 81, 025801

\bibitem[\protect\citeauthoryear{Yakovlev \& Pethick}{Yakovlev \&
  Pethick}{2004}]{yp04}
Yakovlev D.~G.,  Pethick C.,  2004, Ann.Rev.Astron.Astrophys. 42, 169

\bibitem[\protect\citeauthoryear{Yakovlev, Levenfish  \& Shibanov}{Yakovlev
  et~al.}{1999a}]{yls99}
Yakovlev D.~G.,  Levenfish K.~P.,   Shibanov Y.~A.,  1999a, Phys. Usp. 42, 737

\bibitem[\protect\citeauthoryear{Yakovlev, Kaminker  \& Levenfish}{Yakovlev
  et~al.}{1999b}]{YKL}
Yakovlev D.~G.,  Kaminker A.~D.,   Levenfish K.~P.,  1999b, Astron. Astrophys.
  343, 650

\bibitem[\protect\citeauthoryear{Yakovlev, Kaminker, Gnedin  \&
  Haensel}{Yakovlev et~al.}{2001}]{ykgh01}
Yakovlev D.~G.,  Kaminker A.~D.,  Gnedin O.~Y.,   Haensel P.,  2001, Phys.
  Rep., 354, 1

\makeatother
\end{thebibliography}
\end{document}